\documentclass[smallextended]{svjour3}
\smartqed

\usepackage{bm}
\usepackage{amsfonts}
\usepackage{graphicx}

\journalname{General Relativity and Gravitation}

\begin{document}

\title{Dark Energy and Extending the Geodesic Equations of Motion:}
\subtitle{Connecting the Galactic and Cosmological Length Scales}  

\titlerunning{Connecting the Galactic and Cosmological Scales}

\author{A.~D.~Speliotopoulos}

\institute{
Department of Physics
University of California
Berkley, CA 94720, and 
Department of Mathematics,
Golden Gate University,
San Francisco, CA 94105,
ads@berkeley.edu
}

\date{June 30, 2010}

\maketitle

\begin{abstract}

Recently, an extension of the geodesic equations of motion using the
Dark Energy length scale was
proposed. Here, we apply this extension to the analyzing the 
motion of test particles at the galactic scale and longer. A
cosmological check of the extension is made using the 
observed rotational velocity curves and core sizes of 1393 spiral
galaxies. We derive the density profile of a model galaxy using this
extension, and with it, we calculate $\sigma_8$ to be $0.73_{\pm
  0.12}$; this is within experimental error of the WMAP value of 
$0.761_{-0.048}^{+0.049}$. We then calculate $R_{200}$ to be $206_{\pm 53}$
kpc, which is in reasonable agreement with observations. 
 
\end{abstract}


\maketitle


\section{Introduction}

In a previous paper \cite{ADS}, we constructed an extension of the
geodesic equations of motion (GEOM). This construction is possible
because with the discovery of Dark Energy, $\Lambda_{DE} =
\left(7.21^{+0.82}_{-0.84}\right) \times 10^{-30} \hbox{g/cm}^3$
\cite{Ries1998} - \cite{WMAP}, 
there is now a length scale, $\lambda_{DE} =c/(\Lambda_{DE}G)^{1/2}$,
associated with the universe.  As this length scale is also not
associated with the mass of any known particle, this extension does
not violate various statements of the equivalence
principle. Importantly, the extension does not change the GEOM
for massless particles, and thus astronomical 
observations of the universe\textemdash which are based on the
trajectory of photons\textemdash remain unchanged. At
$14010^{800}_{820}$ Mpc, the shear scale of 
$\lambda_{DE}$ ensures that effects of this extension will not have
already been observed either in the motion of bodies in the solar
system, or in terrestrial experiments. Indeed, by analyzing the
effects of the extension at these length scales, we established a
lower bound to $\alpha_{\Lambda}$.  This $\alpha_\Lambda$ is the only
free parameter in the theory, and gives the power law dependence of
the extension on the ratio $c^2 R/\Lambda_{DE}G$, where $R$ is the Ricci
scalar. 

At the conclusion of \cite{ADS}, we argued that it is only at galactic
length scales and longer will we expect effects of the extension
to be relevant.  A study of these effects is the focus of this paper.
In particular, we analyze the restrictions on our extension of the
GEOM due to both observations of galactic structure, and recent
measurements by WMAP. We find that these observations and measurements
do not rule out our extension of the GEOM. To the contrary, by applying
the extension to an analysis of the motion of stars in spiral
galaxies, we are able to calculate $\sigma_8$\textemdash the rms
fluctuation in the density of matter at $8 h^{-1}$ Mpc\textemdash to
be $0.73_{\pm0.12}$. This value is in excellent agreement with the
WMAP measurement of $0.761^{+0.049}_{-0.048}$ for $\sigma_8$. We are
also able to calculate $R_{200}$\textemdash the distance from the
center of a galaxy at which the density of matter equals 200 times
that of the critical 
density\textemdash to be $206_{\pm 53}$ kpc, which is in reasonable agreement
with observations. Calculations of both quantities are possible because
of an unexpected connection between the effects of our extension on
the motion of test particles at galactic scales with cosmological
length scales.  This connection allows us to set a definite value of
$1.56_{\pm 0.10}$ for $\alpha_{\Lambda}$ that is equal to the rough 
lower bound on $\alpha_\Lambda$ established in \cite{ADS} for effects
of the extension to be unobservable at solar system and terrestrial scales.

This stringent test of the extended GEOM is only 
possible because of the scale of $\lambda_{DE}$.  With a value of
$14010^{800}_{820}$ Mpc, it is expected that any effects of the extension
will only be apparent at the galactic scale or 
longer. (Indeed, $\lambda_{DE}$ is so large that it is only because of
the nonlinear dependence of the extension on $R$ that effects at the galactic
scale are apparent at all.) It is also precisely at the galactic length
scale where deviations from motion under Newtonian gravity
appear. Using a simple model of a spiral galaxy, we are able to
determine the density profile of the galaxy by applying our extension
of the GEOM to the motion of stars within it. We find that
effects of our extension necessarily extend beyond the galactic
scale. Not unexpectedly, at these scales the $1/r$ interaction
potential that is expected between galaxies from Newtonian gravity is
now logarithmic, which is consistent with the interaction potential
between galaxies and galactic clusters inferred through
observations. At even larger length scales, we find that the predicted density
profile for the model galaxy goes to zero exponentially fast at distances 
beyond the Hubble length scale, $\lambda_H = c/H$, (where $H = h H_0$,
and $H_0 = 100$ $km/s/$ Mpc) from the center of the model
galaxy. While this result is certainly physically reasonable,
it is surprising that such a length scale naturally appears in
the theory, \textit{even though a cosmological model is not mentioned
  either in its construction, or in its analysis}. This unexpected
connection with cosmology allows us to set the value of $\alpha_\Lambda$ to be
$1.56_{\pm0.10}$ using the WMAP measurements of $h$ and $\Lambda_{DE}$. 

We have also used this density profile to calculate explicitly
$\sigma_8$. This calculation is possible because of 
four data sets in the literature, \cite{Blok-1} - \cite{Math}. These
data sets are the result of observations\textemdash made over a
thirty-year span\textemdash of galactic rotation curves that give both
the asymptotic rotational velocity and the core sizes of 1393 spiral
galaxies. While only 
a very small fraction of the observed galaxies in the universe, the size of
the data set and the fact that the vast majority of these observations
were unbiased, allows us to obtain average values for the parameters
used to characterize the model galaxy.  These parameters, along with the
value of $\alpha_\Lambda$, is used to calculate a
$\sigma_8$ that is in excellent agreement with WMAP
measurements. We are also able to use these parameters to predict a 
value of $206_{\pm 53}$ kpc for $R_{200}$, which is in reasonable agreement
with observations.    

Like Peebles' model for structure formation \cite{PeeblesBook}, the
total density of matter for our model galaxy can be written as a sum
of a asymptotic, background density, $\rho_{\hbox{\scriptsize
    {asymp}}}(\mathbf{x})$, and a linear perturbation,
$\rho_{II}^{1}(\mathbf{x})$. The matter that makes up  
$\rho_{\hbox{\scriptsize asymp}}(\mathbf{x})$ does not contribute to
the motion of stars within the galaxy, while the matter that makes up 
$\rho_{II}^{1}(\mathbf{x})$ does.  Unlike Peebles' model, however,
$\rho_{\hbox{\scriptsize {asymp}}}(\mathbf{x})$ is not a constant, but
instead varies inversely with distance from the galactic core, and
dies off exponentially fast beyond the Hubble scale. Moreover, the
form of $\rho_{\hbox{\scriptsize{asymp}}}(\mathbf{x})$ depends only on the dimensionality of
spacetime and $\alpha_\Lambda$, while the scale at
which it decreases depends only on $\lambda_{DE}$. In this sense,
$\rho_{\hbox{\scriptsize{asymp}}}(\mathbf{x})$ is universal, and does not depend on
the detail structure of the galaxy.  This is in contrast to
$\rho^{1}_{II}(\mathbf{x})$, which depends explicitly on the structure
of the galaxy near the galactic core both in form and in scale.

While $\rho_{\hbox{\scriptsize{asymp}}}(\mathbf{x})$ does not
contribute to the motion of stars within galaxies, it does
contribute to the deflection of light. By construction, the extended
GEOM does not affect the equations of motion for massless
particles. Light still travels along geodesics, and the degree of
the deflection of light is determined by the total local density of matter.
As such, using the deflection of light to
measure the local matter density will result in a measurement of
$\rho_{\hbox{\scriptsize{asymp}}}(\mathbf{x}) +
\rho^{1}_{II}(\mathbf{x})$. In contrast, the motion of
stars in galaxies are affected by our extension of the GEOM.  As such,
their motion is determined solely by $\rho^{1}_{II}(\mathbf{x})$, and
thus when this motion is used to determine the 
local matter density, what is measured is only the matter that makes
up $\rho^{1}_{II}(\mathbf{x})$. As we find $\rho^{1}_{II} <<
\rho_{\hbox{\scriptsize{asymp}}}$ outside of a few galactic core
radii, the presence of the vast majority of matter in the universe can
only be inferred by though the effects that the local density of
matter has on the trajectory of light. 

The rest of the paper is divided into five parts. In the first part, an
overview of the extended GEOM is given, and the properties of the
extension needed in this paper is outlined. In the second part,
we introduce our model of the galaxy, and using the extended GEOM, we
derive the density profile of the galaxy\emph{ given} a rotational
velocity curve for it. We show that this density dies off exponentially
fast beyond a fixed distance from the galaxy.  In the third part, we
use WMAP measurements of $h$ to determine $\alpha_{\Lambda}$, and
calculate $\sigma_8$ and 
$R_{200}$ using the density profile of our model galaxy and
observational data on galactic rotation curves from the 
literature. The value for $\sigma_8$ is then compared with WMAP
measurements.  In the fourth part, we calculate the gravitational
potential for the model galaxy, and determine which portion of the
density can be determined through direct observations of the motion of
stars in the galaxy, and which can only be determined
through the deflection of light.  Concluding remarks can then be found
in the last part.
 
\section{An Overview of the Extended GEOM}

While there is currently no consensus as to the nature of Dark
Energy, modifications to Einstein's equations to include the cosmological 
constant are both well known and minimal. In addition, WMAP
measured the ratio of the pressure to energy density ratio for Dark
Energy to be $-0.967^{+0.073}_{-0.072}$; this is within
experimental error of $-1$, the ratio expected for the cosmological
constant. We thus identify Dark Energy with the cosmological
constant in this paper, and require only  
that $\Lambda_{DE}$ changes so slowly that it can be considered a
constant in our analysis. Einstein's field equations are 
then
\begin{equation}
  R_{\mu\nu} - \frac{1}{2}g_{\mu\nu}R +
  \frac{\Lambda_{DE}G}{c^2} g_{\mu\nu}= - \frac{8\pi G}{c^4} T_{\mu\nu},
\label{EinsteinEquation}
\end{equation}
where $T_{\mu\nu}$ is the energy-momentum tensor for matter,
$R_{\mu\nu}$ is the Ricci tensor, Greek indices run from $0$ to $3$,
and the signature of $g_{\mu\nu}$ is $(1,-1,-1,-1)$. 

The extended GEOM for a test particle with mass, $m$, is obtained from
the Lagrangian
\begin{equation}
\mathcal{L}_{\hbox{\scriptsize{Ext}}} \equiv
\mathfrak{R}[c^2R/\Lambda_{DE}G]
\left(g_{\mu\nu}\dot{x}^\mu\dot{x}^\nu\right)^{\frac{1}{2}}. 
\end{equation} 
While the function $\mathfrak{R}$ is arbitrary, in \cite{ADS} we argued
that the simplest choice is $\mathfrak{R}(x) =
\left[1+\mathfrak{D}(x)\right]^{1/2}$, where
\begin{equation}
\mathfrak{D}(x) = \chi(\alpha_\Lambda)\int_x^\infty
\frac{ds}{1+s^{1+{\alpha_\Lambda}}},
\end{equation}
while 
\begin{equation}
\frac{1}{\chi(\alpha_\Lambda)}\equiv \int_0^\infty
\frac{ds}{1+s^{1+{\alpha_\Lambda}}}
=\frac{\sin\left[\pi/(1+\alpha_{\Lambda})\right]}{\pi/(1+\alpha_{\Lambda})}, 
\end{equation}
is defined so that $D(0)=1$. Here, $\alpha_{\Lambda}$ is a constant, and is the
only free parameter in the theory. In \cite{ADS}, we showed that for
the effects of the extension not to have already been observed in
terrestrial experiments, $\alpha_\Lambda$ must be between $1.28$
(for $\Lambda_{DE} = 10^{-32}$ g/cm${}^3$) and 1.58 (for $\Lambda_{DE}
= 10^{-29}$ g/cm${}^3$).  

While the equations of motion derived from
$\mathcal{L}_{\hbox{\scriptsize EXT}}$ is 
\begin{equation}
\frac{D^2 x^\mu}{\partial t^2} = c^2 \left(g^{\mu\nu} - \frac{v^\mu
  v^\nu}{c^2}\right)
  \nabla_\nu\log\mathfrak{R}[c^2R/\Lambda_{DE}G],
\label{genEOM}
\end{equation}
we are interested in the motion of stars in galaxies. Moreover,
WMAP and the Supernova Legacy Survey 
put $\Omega_K = -0.011_{\pm 0.012}$, and thus the universe is
essentially flat.  Indeed, WMAP's value for $h$ is determined with
this assumption. As such, we 
are working in the nonrelativistic and weak gravity limits, and
therefore take the metric to be $g_{\mu\nu} =
\eta_{\mu\nu}+h_{\mu\nu}$.  Here, $\eta_{\mu\nu}$ is the Minkowski
metric, and $h_{\mu\nu}$ is a small perturbation. The only
nonzero component of $h_{\mu\nu}$ is $h_{00}=2\Phi/c^2$, where $\Phi$
is the Newtonian gravitational
potential. Equation $(\ref{EinsteinEquation})$ then reduces to 
\begin{equation}
\mathbf{\nabla}^2\Phi + 2\frac{\Lambda_{DE}G}{c^2}\Phi = 4\pi\rho G -
\Lambda_{DE}G, 
\label{PhiEOM}
\end{equation}
with the additional terms due to the cosmological
constant. 

As we are dealing with the motion of stars in galaxies,
taking $\eta_{\mu\nu}$ to be the background metric would seem to be
straightforward. There is one subtlety, however. Even when
$T_{\mu\nu}=0$ in Eq.~$(\ref{EinsteinEquation})$, 
$\Lambda_{DE}$ is present, and the spacetime is not flat; at scales
comparable to $\lambda_{DE}$, the spacetime will be significantly
different from Minkowski space even in the absence of matter. At
$14010^{800}_{820}$ Mpc, $\lambda_{DE}$ is over three times the Hubble 
scale, $\lambda_H$, however, and taking the background metric to be
flat is a good approximation throughout most of the physically relevant
length scales. By restricting ourselves to length
scales much less than $\lambda_{DE}$, taking $\eta_{\mu\nu}$ as the
background metric is a good approximation, and the terms proportional
to $\Lambda_{DE}$ in Eq.~$(\ref{PhiEOM})$ can be neglected. 

Using Eq.~$(\ref{PhiEOM})$, Eq.~$(\ref{genEOM})$ reduces to   
\begin{equation}
\frac{d^2 \mathbf{x}}{dt^2} \approx -\mathbf{\nabla} \Phi + \left(\frac{4\pi
     c^2\chi}{\Lambda_{DE}}\right)
     \left\{1+\left(4+\frac{8\pi\rho}{\Lambda_{DE}}\right)^{1+{\alpha_\Lambda}}\right\}^{-1}  
     \mathbf{\nabla}\rho,
\label{FinalEOM}
\end{equation}
in the nonrelativistic and weak gravity limits. 
Here, we have related $R$ to the trace of the energy-momentum tensor,
$T$, using Eq.~$(\ref{EinsteinEquation})$. We have also assumed that the
spacetime is spatially symmetric, and that the particle moves through
an ambient, nonrelativistic fluid with density, $\rho$. We are dealing
with only gravitational forces, and thus do not differentiate between
baryonic matter and Dark Matter in $\rho$. As shown in 
\cite{ADS}, the energy-momentum tensor for the fluid can be
approximated as $T_{\mu\nu}\approx \rho u_\mu u_\nu$ in the nonrelativistic
and weak gravity limits even though elements in the fluid propagate
under the extended GEOM instead of the GEOM. Because
$\Lambda_{DE} \sim 10^{-30}$ g/cm${}^3$, we expect that
$8\pi\rho/\Lambda_{DE}\ge 0$ in galaxies, and thus have also used the
expansion     
\begin{equation}
\mathfrak{D}(4+8\pi\rho/\Lambda_{DE}) = \chi\sum_{n=0}^\infty
  \frac{(-1)^n}{n(1+{\alpha_\Lambda})+{\alpha_\Lambda}}
  \left(4+\frac{8\pi\rho}{\Lambda_{DE}}\right)^{-n(1+{\alpha_\Lambda})-{\alpha_\Lambda}},
\label{D-expand}
\end{equation}
in obtaining Eq.~$(\ref{FinalEOM})$.

\section{Effects on the Galactic Scale}

While definitive, a first principle calculation of the galactic
rotation curves using the extended GEOM to describe the 
motion each star in a galaxy is analytically intractable. 
Instead, \emph{given} a rotation curve for a model galaxy, we
will use the extended GEOM to derive the density profile for the
galaxy, and through this profile, describe the motion of test particles within
it. Not surprisingly, we find that inside the galactic core stars move as
though they are in a Newtonian potential.
Outside of the core, on the other hand, stars move as though they are
in a non-Newtonian, logarithmic potential. 

Since the lower bound on $\alpha_{\Lambda}$ is between 1.28 and
1.58, we take $\alpha_{\Lambda}=3/2$ as a guide when making
approximations in this section.  In the next section, we will use WMAP
measurements of $h$ and $\Omega_{\Lambda}$ to determine
$\alpha_\Lambda$.  

\subsection{The Model Galaxy}

A number of geometries have been used to model the formation of
galaxies \cite{Fill} - \cite{Hoff1988}, including a 
spherical geometry. Because we will be making connection with 
cosmology, we are interested in the large-distance properties of the
density profile, and at such distances, detail structures of
galaxies are washed out; only the spherically symmetric features
survive. We thus use a spherical geometry to model our
idealized galaxy, and will focus on the general structure of the
galaxy instead of on the details.  

To determine the density profile, we divide space into the following three
regions. The first, Region I $=\{r \>\>\vert \>\>  r\le r_H
\hbox{, and  }\rho \gg \Lambda_{DE}/2\pi\}$, encompasses the
galactic core with radius, $r_H$. Here, the density of matter is much
larger than $\Lambda_{DE}/2\pi$. The second, Region II $=\{r
\>\>\vert \>\>\> r_H< r \le r_{II}, \hbox{ and  } \rho \gg
\Lambda_{DE}/2\pi\}$, encompasses the region outside the
core that contains stars undergoing rotational motion with constant  
velocity. Here, the density of matter is also greater than
$\Lambda_{DE}/2\pi$, and the region extends out to a distance of
$r_{II}$, which is determined by the theory. While this region of constant
rotational motion is much larger than the size of the
galaxy, there are no other galaxies in the model, and we find that there
is a cutoff in the density at distances beyond $r_{II}$. Finally, the third, Region III
$=\{r \>\>\vert \>\> r_{II} < r \hbox{, and } \rho \ll
\Lambda_{DE}/2\pi\}$, encompasses the region where the density
of matter is less than $\Lambda_{DE}/2\pi$.  We will find that the
density is exponentially small here, and we will see in the next section
that this region is not physically relevant. 

We assume that all the stars in the model galaxy undergo circular
motion. While this is an approximation, galactic rotation curves are
determined with stars that undergo this motion, and we use these
curves as inputs for our analysis. With this approximation, the
acceleration of each star, $\mathbf{a} \equiv \ddot{\mathbf{x}}$, is a
function of $r$ only. We can then take the divergence of
Eq.~$(\ref{FinalEOM})$, and obtain   
\begin{equation}
f(r)= \rho - \chi\lambda_{DE}^2
  \frac{  \left\{\mathbf{\nabla}^2\rho -
  \frac{1+{\alpha_\Lambda}}{4+8\pi\rho/\Lambda_{DE}}
  \left(\frac{8\pi}{\Lambda_{DE}}\right) 
  \vert\mathbf{\nabla}\rho\vert^2\right\}}{\left\{1+\left(4+ 
  \frac{8\pi\rho}{\Lambda_{DE}}\right)^{1+{\alpha_\Lambda}}\right\}},
\label{rhoEOM}
\end{equation}
where $f(r) \equiv -\mathbf{\nabla}\cdot\mathbf{a}/4\pi G$ is
considered a source. Because 
\begin{equation}
-\mathbf{\nabla}\cdot \mathbf{a}
=\frac{1}{r^2}\frac{\partial\>\>}{\partial r}\left[rv(r)^2\right],
\label{div-accel}
\end{equation}
given a velocity curve for the galaxy, $v(r)$, $\mathbf{a}$ can be
found, and the source determined.  

While a number of different models have been used in the literature to
fit the observed rotational velocity curves, we use a
particularly simple idealization of the curves:
\[v^{\hbox{\scriptsize ideal}}(r) = \left\{ 
\begin{array}{l l}
  v_H r/r_H & \quad \mbox{for $r \le r_H$}\\
  v_H & \quad \mbox{for $r\ge r_H$, }\\ 
\end{array} 
\right. 
\]
where $v_H$ is the asymptotic velocity of the velocity
curve. We make this idealization because our purpose
is not to derive the precise density profile of a galaxy; our choice
of geometry does not allow for such a derivation.  Instead, our
purpose is to determine the overall features of the density profile;
for this purpose, $v^{\hbox{\scriptsize{ideal}}}(r)$ is sufficient. 

While $v^{\hbox{\scriptsize ideal}}(r)$ is continuous, $f(r)$ is not. We find 
that 
\[f(r)= \left\{ 
\begin{array}{l l}
  \rho_H =3v_H^2/4\pi G r_H^2 & \quad \mbox{for $r \le r_H$}\\
  \rho_H r_H^2/3r^2 & \quad \mbox{for $r\ge r_H$.}\\
\end{array} 
\right.
\] 
Here, $\rho_H$ is identified as the density of matter in the galactic
core, which is a constant for our idealized velocity curve. 

\subsection{The Density Profile of the Model Galaxy}

In both Regions I and II, $\rho\gg \Lambda_{DE}/2\pi$, and
Eq.~$(\ref{rhoEOM})$ may be approximated as 
\begin{equation}
f(u) = \rho + \frac{\Lambda_{DE}}{8\pi\alpha_\Lambda}
\mathbf{\nabla}_{u}^2\left(\frac{\Lambda_{DE}}{8\pi\rho}\right)^{\alpha_\Lambda}, 
\label{rhoI-II-EOM}
\end{equation}
where ${u} = r/\chi^{1/2}\lambda_{DE}$, and $\mathbf{\nabla}_{u}$
denotes the derivative with respect to ${u}$. 

In Region III, on the other hand, we find that $f(r)$ is
negligibly small, and can be set to zero. Equation $(\ref{rhoEOM})$
then reduces to 
\begin{equation}
0 = \rho -
\frac{1}{1+4^{1+{\alpha_\Lambda}}}\mathbf{\nabla}^2_{u}\rho. 
\label{rhoIII-EOM}
\end{equation}

\subsubsection{The Solution in Regions I and II}

Because $f(u) =\rho_H$ in Region I, we find that the density,
$\rho_I(r)$, in this region is simply $\rho_H$.  

To find the density, $\rho_{II}(r)$, in Region II, asymptotic
analysis \cite{Bender} is used. We first find a solution to
Eq.~$(\ref{rhoI-II-EOM})$ in a region where $\rho_H (r_H/r)^2/3 \ll
\rho(r)$. This is equivalent to making the anzatz   
that within Region II there is a point $r_{\hbox{\scriptsize{asymp}}}$
beyond which the driving
term in Eq.~$(\ref{rhoI-II-EOM})$ is negligibly small compared to the
density. Equation $(\ref{rhoI-II-EOM})$ then reduces to
\begin{equation}
0 = \rho_{\hbox{\scriptsize{asymp}}} 
 + \frac{\Lambda_{DE}}{8\pi\alpha_\Lambda}
\mathbf{\nabla}_{u}^2
\left(
	\frac{\Lambda_{DE}}{8\pi\rho_{\hbox{\scriptsize{asymp}}}}
\right)^{\alpha_\Lambda},
\label{asympEOM}
\end{equation}
where $\rho_{\hbox{\scriptsize{asymp}}}$ is the asymptotic density. 
We then perturb off $\rho_{\hbox{\scriptsize{asymp}}}$ 
to find the density near the galactic core.

The solution to Eq.~$(\ref{asympEOM})$ has the form 
\begin{equation}
\rho_{\hbox{\scriptsize{asymp}}} = \frac{\Lambda_{DE}}{8\pi}
\Sigma({\alpha_\Lambda}) {u}^p,
\end{equation}
where $p$ and $\Sigma(\alpha_\Lambda)$ are solutions to
\begin{equation}
0=1- \frac{p(1-{\alpha_\Lambda}
  p)}{\bigg[\Sigma({\alpha_\Lambda})\bigg]^{1+{\alpha_\Lambda}}} 
\frac{1}{u^{p(1+{\alpha_\Lambda})+2}}.
\label{asympConstraint}
\end{equation}
Thus $p=-2/(1+{\alpha_\Lambda})$, while
$\Sigma(\alpha_\Lambda)$ is given by
\begin{equation}
0=1+\frac{2(1+3{\alpha_\Lambda})}{(1+{\alpha_\Lambda})^2}
\frac{1}{\big[\Sigma({\alpha_\Lambda})\big]^{1+{\alpha_\Lambda}}}. 
\label{amp}
\end{equation}
For $\rho_{\hbox{\scriptsize{asymp}}}$ to be positive, 
$\Sigma({\alpha_\Lambda}) >0$, and thus there must be positive
solutions to Eq.~$(\ref{amp})$. Such solutions exist only if 
${\alpha_\Lambda}$ is the ratio of a odd integer to an even
integer. Our choice of ${\alpha_\Lambda} = 3/2$ satisfies this
criteria, and we arrive at the following asymptotic solution
\begin{equation}
\rho_{\hbox{\scriptsize{asymp}}}(r) = \frac{\Lambda_{DE}}{8\pi} \Sigma({\alpha_\Lambda})
\left(\frac{\chi\lambda_{DE}^2}{r^2}\right)^{1/(1+{\alpha_\Lambda})},
\label{asymp}
\end{equation}
with
\begin{equation}
\Sigma({\alpha_\Lambda}) =
\left[\frac{2(1+3{\alpha_\Lambda})}{(1+{\alpha_\Lambda})^2}\right]^{1/(1+{\alpha_\Lambda})}.
\label{Sigma}
\end{equation}

To justify our anzatz that $r_{\hbox{\scriptsize{asymp}}}$ lies within
Region II, we set $f(r_{\hbox{\scriptsize{asymp}}})=
\rho_{\hbox{\scriptsize{asymp}}}(r_{\hbox{\scriptsize{asymp}}})$, and
find the bound
\begin{equation}
\frac{r_{\hbox{\scriptsize{asymp}}}}{r_H} =
\left(\frac{8\pi\rho_H}{3\Lambda_{DE}}
\frac{u_H^{2/(1+\alpha_\Lambda)}}
     {\Sigma(\alpha_\Lambda)}\right)^{(1+\alpha_\Lambda)/2\alpha_\Lambda},
\label{RII}
\end{equation}
where $u_H = r_H/\chi^{1/2}\lambda_{DE}$. For ${\alpha_\Lambda} =
3/2$, $\rho_H\sim 10^{-24}$ g/cm${}^3$, and $r_H=10$ kpc,
$r_{\hbox{\scriptsize{asymp}}}\le 6.8 r_H$; the anzatz is   
thus valid throughout the great majority of Region II for the range of
galaxies we are interested in. The upper limit, $r_{II}$, to Region II,
on the other hand, is found by setting 
$8\pi\rho_{\hbox{\scriptsize{asymp}}}(r_{II})/\Lambda_{DE}=4$. This gives
\begin{equation}
r_{II} =
\left[\frac{1}{4}\Sigma({\alpha_\Lambda})\right]^{(1+{\alpha_\Lambda})/2}
\chi^{1/2}\lambda_{DE}. 
\label{firstRII}
\end{equation}
For ${\alpha_\Lambda} = 3/2$, $r_{II} \approx 0.27\>\lambda_{DE}$.

Next, to see the structural details of the galaxy, we take $\rho_{II} =
\rho_{\hbox{\scriptsize{asymp}}}+\rho_{II}^{1}$. Expanding
Eq.~$(\ref{rhoI-II-EOM})$ to first order in $\rho_{II}^{1}$ gives 
\begin{equation}
\frac{2(1+3{\alpha_\Lambda})}{(1+{\alpha_\Lambda})^2}
\left[\frac{\rho_H}{3}\left(\frac{{u}_H}{{u}}\right)^2 \right]=
\frac{2(1+3{\alpha_\Lambda})}{(1+\alpha_\Lambda)^2}
\frac{\widehat{\rho}_{II}^{\>1}}{{u}^2} +
\mathbf{\nabla}^2_{u} \widehat{\rho}^{\>1}_{II}.
\label{rho1-EOM}
\end{equation}
where $\widehat{\rho}^{\>1}_{II} ={u}^2\rho^{1}_{II} $. The particular
solution to Eq.~$(\ref{rho1-EOM})$ is again the constant solution, but
now for $\widehat{\rho}^{\>1}_{II}$; this corresponds to
$\rho^{1}_{II} = \rho_H r_H^2/3r^2$. The solution of the homogeneous
equation is straightforward, and combined with
$\rho_{\hbox{\scriptsize{asymp}}}$ from Eq.~$(\ref{asymp})$, we
arrive at the density profile in Region II
\begin{eqnarray}
\rho_{II}(r) = &{}&\frac{\Lambda_{DE}}{8\pi} \Sigma({\alpha_\Lambda})
\left(\frac{\chi\lambda_{DE}^2}{r^2}\right)^{1/(1+{\alpha_\Lambda})} +
\frac{1}{3}
\rho_H\left(\frac{r_H}{r}\right)^2 + 
\nonumber
\\
&{}&\left(\frac{r_H}{r}\right)^{5/2}
\left(C_{\cos}\cos\left[\nu\log\left(\frac{r}{r_H}\right)\right] +
C_{\sin}\sin\left[\nu\log\left(\frac{r}{r_H}\right)\right]\right).
\label{rho-beta}
\end{eqnarray}
Here $\nu = \left[2(1+3\alpha_\Lambda)/(1+\alpha_\Lambda)^2 -
  1/4\right]^{1/2}$, while  
\begin{eqnarray}
C_{\cos} &=& \frac{2}{3}\rho_H
-\rho_{\hbox{\scriptsize{asymp}}}(r_H),
\nonumber
\\
\nu C_{\sin} &=& \frac{7}{3}\rho_H
-\frac{1}{2}\frac{(1+5\alpha_\Lambda)}{(1+\alpha_\Lambda)}
\rho_{\hbox{\scriptsize{asymp}}}(r_H),
\end{eqnarray}
are determined by the boundary conditions for $\rho(r)$ at $r_H$.

The density, $\rho_{II}(r)$, thus consists of the sum of two parts. The
first part, $\rho_{\hbox{\scriptsize{asymp}}}(r)$, corresponds to a
background, asymptotic density, and depends solely on Dark Energy,
fundamental constants, the exponent $\alpha_\Lambda$, and the
dimensionality of spacetime. \textit{It is 
universal, and has the same form irrespective of the detailed
  structure of the galaxy}. The second part, $\rho_{II}^{1}(r)$, on
the other hand, does depend on the detail structure of the 
galaxy. Variations in $\rho_{II}^{1}$ are on the scale of $r_H$, the
core size; in contrast, variations in
$\rho_{\hbox{\scriptsize{asymp}}}$ are on the scale of
$\lambda_{DE}$, the Dark Energy 
length scale. While our analysis is done only to first order in the 
perturbation of $\rho_{II}$, these features of $\rho_{II}^{1}$ hold
to higher orders as well. 

The perturbation, $\rho_{II}^{1}$, itself depends on two terms. The first
has a power law dependence of $(r_H/r)^2$, while the second has a
power law dependence of $(r_H/r)^{5/2}$. Thus, near the galactic
core $\rho_{II}^{1} \sim 1/r^{5/2}$, while for $r\gg r_H$, $\rho_{II}^{1}\sim
1/r^2$. This behavior at large $r$ is consistent with the
pseudo-isothermal density profile \cite{Blok-1} observed in spiral
galaxies. Both terms in the perturbation decrease rapidly with $r$,
and thus detail structural features of the galaxy disappear
quickly with distance from its core.    

\subsubsection{A Natural Cutoff for $\rho(r)$}

With the boundary condition $\rho_{II}(r_{II}) =
\rho_{III}(r_{II})$, the only solution to Eq.~$(\ref{rhoIII-EOM})$ in
Region III that is spherically symmetric and finite as $r\to\infty$ is
\begin{eqnarray}
\rho_{III}(r) = &{}&\frac{\Lambda_{DE}}{8\pi}\Sigma({\alpha_\Lambda})
\frac{\sqrt{\chi}\lambda_{DE}}{r}
\left(1+4^{1+{\alpha_\Lambda}}\right)^{\frac{1}{2}\frac{(1-{\alpha_\Lambda})}{(1+{\alpha_\Lambda})}}
\nonumber
\\
&{}&
\exp\left(1-\frac{r}{\lambda_{DE}}
\sqrt{
     \frac{1+4^{1+{\alpha_\Lambda}}}{\chi}
}\right).
\label{rho3-solution}
\end{eqnarray}
In this region, the density decreases to zero exponentially fast, and
the profile essentially cuts off at
$\sqrt{\chi/(1+4^{1+\alpha_{\Lambda}})}\lambda_{DE}$; there is
therefore a natural cutoff built within the theory. The scale of this
cutoff is approximately $0.20 \>\lambda_{DE}$ for ${\alpha_\Lambda} = 3/2$,
while from Eq.~$(\ref{firstRII})$, $r_{II} \approx 
0.27\>\lambda_{DE}$; to a good approximation $r_{II} \approx
\sqrt{\chi/(1+4^{1+{\alpha_\Lambda}})}\lambda_{DE}$. Regions II and
III thus overlap, and the asymptotic analysis is self-consistent.  As
$\sqrt{\chi/(1+4^{1+{\alpha_\Lambda}}})\lambda_{DE}$ is a scale set by
the theory as opposed to Eq.~$(\ref{firstRII})$, which is set by an
approximation scheme, we take $r_{II} \equiv
\sqrt{\chi/(1+4^{1+{\alpha_\Lambda}})}\lambda_{DE}$
from now on. 

\section{A Cosmological Check}

In this section, we use the recent measurements of the Hubble
constant, $h$, by WMAP to set $\alpha_\Lambda$. This $\alpha_\Lambda$
is then used to calculate $\sigma_8$ and $R_{200}$ using the
model galaxy derived in the previous section. These
calculated values are then compared with WMAP measurements for
$\sigma_8$, and observations of $R_{200}$, thereby providing a
cosmological check of 
the extended GEOM.  

The density profile for the model galaxy in the previous section was
calculated using $\eta_{\mu\nu}$ as the background metric in the
nonrelativistic and weak gravity limits.  While this
approximation is appropriate at the galactic scale, we would
not, in general, expect our $\rho(\mathbf{x})$ to be valid at 
cosmological scales; at one point, we would expect the local curvature
of the universe to introduce corrections that were not taken into
account in the original linearization of the metric,
$g_{\mu\nu}$. WMAP and the Supernova Legacy Survey 
put $\Omega_K = -0.011_{\pm 0.012}$, however, while WMAP and the HST
key project set $\Omega_K = -0.014_{\pm0.017}$
\cite{WMAP}. Measurements have thus shown that the universe is within 
experimental error of being flat, and indeed, WMAP's determination of
$h=0.732_{-0.032}^{+0.031}$ was made under this assumption. As the
density profiles of galaxies do not increase at large distances from
the galactic core, it is consistent\textemdash both internal to the
perturbative analysis and external with experimental
observations\textemdash to extend our solution in
Eq.~$(\ref{rho-beta})$ to cosmological scales. 

Like cosmological models, the spacetime considered in Sec 2 is divided
into time-slices. Also like cosmological models, the radial 
coordinate, $r$, used in this paper is the \emph{proper distance}
between two points on the same time-slice. This distance is not what is
observed in astronomical observations, however. An object separated
from an observer by a proper-distance can only be observable some time
in the future, when the object enters the past 
lightcone of the observer. This is an important distinction that will
be used when setting $\alpha_\Lambda$.  

Our model from Sec 2 is for a single galaxy, while the universe
certainly contains more than just one galaxy. That we can apply our
single-galaxy model to observations made on the universe such as those
by WMAP is due to the large separations between galaxies, and the
universal properties of
$\rho_{\hbox{\scriptsize{asymp}}}$. Separations between galaxies are 
observed to be on the mega-parsec scale, while the core size of
galaxies, $r_H$, are on the kilo-parsec scale.  As we showed in Sec 2, the
detailed structure of the model galaxy is described by
$\rho_{II}^{1}$, which dies off as $1/r^2$ for large $r$. Much beyond
$\sim6.8 r_H$, knowledge of the detail structure of galaxies is washed
out, and only 
$\rho_{\hbox{\scriptsize{asymp}}}$ remains.  Thus, the
single galaxy model in Sec 2 is an adequate model of a galaxy in
the universe when it is limited to some region centered on the
galaxy, and because of the universal properties of
$\rho_{\hbox{\scriptsize{asymp}}}$, the extent of this region can be
quite large. Indeed, due to the large separation between galaxies, it
would be appropriate to extend this region to the surface of a sphere
of radius, $d$, where $2d$ is the proper distance from the galaxy to
its nearest neighbor galaxy. (If the density is required to be
everywhere continuous, the shape of this region can be adjusted to
ensure that this requirement is met.) The profile of our model galaxy
is applicable within this sphere. Moreover, we are able to describe a
distribution of many galaxies in the universe by replicating the
single-galaxy model; the location and parameters of the replicated
model need only be changed to reflect the properties of each galaxy in
the distribution.  

\subsection{$\alpha_\Lambda$ and the Hubble Length, $\lambda_H$}

The density profile for the model galaxy in Sec 2 decreases
exponentially to zero beyond a distance $r_{II}$ from the center of the
galaxy. As $r_{II}$ does not depend on the detailed structure of
galaxies, and because a precipitous decrease in the local density of
matter has not been observed in the 
universe, we argue below that we must have $d < r_{II}$ between any
two galaxies in our distribution of galaxies. This 
restriction determines $r_{II}$, which in turn will determine
$\alpha_\Lambda$.

Consider a two spiral galaxy model where the two galaxies are
separated by a distance $2d$; galaxy A is located at $\mathbf{d}$,
while galaxy B is located at $-\mathbf{d}$.  We require that $d$ be
much larger than the size of either galaxy. We also
require that while the two galaxies may move relative to each other,
this motion takes place over a very long time scale, and thus
$\mathbf{d}$ can approximated as a constant independent of time. This
is a restrictive requirement; observations show that galaxies do orbit
one another, and there can be significant relative motion of the
galaxies. We do not need to construct a detail model of the dynamics
of galaxies\textemdash or even the density profile of them\textemdash
to set a value for $\alpha_\Lambda$, however. All that is required is to 
determine the maximum separation between galaxies within the theory,
and then to compare this length with the proper-distance between 
galaxies in cosmological models.  This can be done within a model where
the two galaxies do not move relative to one another.

As with the single galaxy case, we approximate the motion of stars in
both galaxies as undergoing circular motion about the center of each
galaxy. Since the separation between galaxies is
much larger than the size of galaxies, the motion of stars within one
galaxy does not affect the motion of stars within the other. Thus,
Eq.~$(\ref{rhoI-II-EOM})$ still holds 
near the core of galaxy A after replacing
$f(\mathbf{x})$ by $f^{A}(\mathbf{x} - \mathbf{d})$. It also holds near the
core of galaxy B after replacing
$f(\mathbf{x})$ by $f^{B}(\mathbf{x}+\mathbf{d})$. Here, $f^{A}$ 
and $f^{B}$ have the same form as $f(\mathbf{x})$ given by
$v^{\hbox{\scriptsize{ideal}}}(r)$, but with $r$ replaced by
$\vert\mathbf{x}\pm\mathbf{d}\vert$, and with $r_H$ and $v_H$ replaced
by the core size and asymptotic rotational velocity of the respective
galaxy.   

While we can, at least in principle, follow the same analysis as
before to determine the density profile of the galaxies\textemdash the
core of the two galaxies are well separated from one
another\textemdash this detailed analysis is not needed to determine
$\alpha_\Lambda$. Instead, we note that as in the single
galaxy case, spacetime can be divided into two domains,
$\mathfrak{D}$ and $\mathfrak{Z}$. In the first, $\mathfrak{D}$, $\rho
>>\Lambda_{DE}/2\pi$, and asymptotic analysis can still be used to
determine the density profile of the galaxies as
$\rho(\mathbf{x})=
\rho_{\hbox{\scriptsize{asymp}}}(\mathbf{x})+
\rho_{II}^{1A}(\mathbf{x}-\mathbf{d}) +
\rho_{II}^{1B}(\mathbf{x}+\mathbf{d})
$. 
Here, $\rho_{\hbox{\scriptsize{asymp}}}$ is a solution of
Eq~$(\ref{asympEOM})$, but now in the presence of two galaxies instead
of one. (While the presence of the galaxies does not explicitly appear
in this equation, the locations of the galaxies determine the boundary
conditions for $\rho$, and thus establish the underlying geometry as
being axially symmetric.) As before,
$\rho_{II}^{1A}(\mathbf{x}-\mathbf{d})$ and 
$\rho_{II}^{1B}(\mathbf{x}+\mathbf{d})$ are the first-order
perturbations off $\rho_{\hbox{\scriptsize{asymp}}}(\mathbf{x})$ near the core of
each galaxy; they establish the structure of each. Because of the
large separation between galaxies, $\rho_{II}^{1A}$ and
$\rho_{II}^{1B}$ can be solved independently using the equivalent of
Eq.~$(\ref{rho1-EOM})$ for the particular
$\rho_{\hbox{\scriptsize{asymp}}}$ obtained for the two galaxy model, and 
with the origin shifted to the center of each galaxy. 

In the second, $\mathfrak{Z}$, $\rho<<\Lambda_{DE}/2\pi$, and in this
domain, Eq.~$(\ref{rhoIII-EOM})$ holds.  As such, the density
decreases exponentially to zero, and for most of the region, we may take $\rho
\approx 0$. Note that this behavior of $\rho$ is set solely by the
form of Eq.~$(\ref{rhoIII-EOM})$, and does not depend on either the
distribution of the galaxies, or the detail structure of them.

Consider now the behavior of the density profile along the line
connecting the cores of the two galaxies.  If the separation, $2d$, between
the two galaxies is smaller than some distance $2d_{\hbox{\scriptsize{max}}}$,
then for all points along this line $\rho > \Lambda_{DE}/2\pi$. The
density, $\rho(\mathbf{x})$, is locally maximum at the core,
 $\mathbf{x}=-\mathbf{d}$, of galaxy B, and decreases as
the distance, $\vert\mathbf{x}+\mathbf{d}\vert$, from the core
increases. Since Eq.~$(\ref{asympEOM})$ does not depend on the
detail structure of either galaxy, the density reaches a local minimum, 
$\rho_{\hbox{\scriptsize{min}}}\equiv
\rho(0)$, at the midpoint between the two galaxies. The density then
increases as the distance increases until it is again a local maximum
at the core, $\mathbf{x}=\mathbf{d}$, of galaxy A.  When
$d<d_{\hbox{\scriptsize{max}}}$, 
$\rho_{\hbox{\scriptsize{min}}}>\Lambda_{DE}/2\pi$, and in this case 
$\mathfrak{D}$ is simply connected. 

To estimate the scale of $d_{\hbox{\scriptsize{max}}}$, consider the
limit of large $d$.  As the separation between galaxies increases,
$\rho_{\hbox{\scriptsize{min}}}$ decreases until at one point
$\rho_{\hbox{\scriptsize{min}}}<<\Lambda_{DE}/2\pi$. When this
happens, the midpoint between the galaxies is contained in $\mathfrak{Z}$, and
$\mathfrak{D}$ is no longer simply connected. Instead, $\mathfrak{D} =
\mathfrak{D}_A \cup \mathfrak{D}_B$, where $\mathfrak{D}_A$ and
$\mathfrak{D}_B$ contains the core of galaxy A and galaxy B,
respectively, and $\mathfrak{D}_A\cap 
\mathfrak{D}_B = \emptyset$.  The two galaxies are separated by a
region in $\mathfrak{Z}$ where $\rho$ decreases exponentially fast. As
such, for large enough separations the density  
profile of one galaxy will not depend on the presence of
the other, and we will be back to the single galaxy case. For
each galaxy, $\rho$ is now given by Eq.~$(\ref{rho-beta})$, with $r$
measured from the center of the galaxy, and the $r_H$ and
$v_H$ set by the properties of this galaxy.  In this limit,
$\mathfrak{D}_A$ and 
$\mathfrak{D}_B$ are spheres centered at the core of their respective
galaxy, and each has radii $r_{II}$. Since $2d_{\hbox{\scriptsize{max}}}$
is the maximum separation between galaxies for which $\mathfrak{D}$ is
simply connected, we estimate $d_{\hbox{\scriptsize{max}}} = r_{II}$.   

As astronomical observations have not shown that the density of matter
decreases precipitously at any point along the separation two
galaxies, we require that in our model of a distribution of galaxies,
the separation between galaxies must be less than $2r_{II}$.

Since $2r_{II}$ is the maximum possible proper distance between
galaxies in the model, it must be cosmological in scale. It
would be physically appropriate to identify this scale with
$\lambda_{eH}$\textemdash the proper distance to the cosmological
event horizon at present day\textemdash for the following
reasons. Consider an observer on galaxy A, which at the present day,
$t_0$, is separated from galaxy B by a proper distance $r_{II}$. If $2r_{II} <
\lambda_{eH}$, at a finite time, $t$, in the future galaxy B will
pass into the past lightcone of galaxy A, and thus be observed by the
observer on A. If, on the other hand, $2r_{II}>\lambda_{eH}$, then
for no time $t$ will galaxy B pass within the pass lightcone of galaxy
A, and will never be observed by the observer on A. More importantly,
when $2r_{II} > \lambda_{eH}$, no physical property of galaxy A can be
influenced by galaxy B in the future, and thus the presence of galaxy
B has no physical significance to the observer on A. Galaxy A and
galaxy B will be causally disconnected.

While these are good, physical reasons for identifying $2r_{II}$ with
$\lambda_{eH}$, it is not physically possible to measure
$\lambda_{eH}$  directly. All  
physical measurements are made on objects within or on the past
lightcone of an observer. Because $\lambda_{eH}$ is the proper
distance to the event horizon at present day, it cannot, even in 
principle, be measured; it is at best inferred. What is physically
measurable, on the other hand, is $\lambda_H$, which is also the
radius of the Hubble sphere.  Although it 
is shown in \cite{Davis2004} that for a $\Lambda$CDM model with
$(\Omega_M,\Omega_\Lambda) = (0.3,0.7)$ and $h=0.70$ the 
cosmological event horizon is not equivalent to the Hubble sphere,
$1/H$ is the time it has taken galaxies to achieve their current
separations \cite{PeeblesBook}. The Hubble scale is the maximum
distance that light has 
traveled in this time. Thus, $\lambda_H$ is the maximum separation
between galaxy A and galaxy B from which an observer on one could have
observed the formation of the other at some time in the past; it would
be possible for the formation of one galaxy can influence the
formation of the other. 

There are thus also good physical reasons for identifying $2r_{II}$ with
$\lambda_H$. As the difference between $\lambda_{eH}$ and $\lambda_H$
calculated in \cite{Davis2004} is very small at present day, and because 
$\lambda_H$ is a natural length scale in cosmology that is explicitly
measurable, we identified $2r_H = \lambda_H$. Since we can express
\begin{equation}
r_{II} =
\left[\frac{8\pi\chi}{3\Omega_\Lambda(1+4^{1+\alpha_\Lambda})}\right]^{1/2}
\lambda_H, 
\label{r-II-Lambda}
\end{equation}
this identification sets the value for $\alpha_{\Lambda}$. Here,
$\Omega_\Lambda = 0.716_{\pm 0.055}$ from \cite{WMAP} is the 
fractional density of Dark Energy in the universe. The solution to
Eq.~$(\ref{r-II-Lambda})$ gives $\alpha_\Lambda = 1.56_{\pm 0.10}$,
which is close to the working value of $3/2$ for
$\alpha_\Lambda$ used in the previous section. Note that to the
lowest significant figure, this $\alpha_\Lambda$ is the ratio of an
odd to an even integer as well.

\subsection{$\sigma_8$, $R_{200}$, and Observational Data for the
  Galactic Rotation Curves}  

Following \cite{TurnerBook} and \cite{PeeblesBook}, linear
fluctuations in the density within a sphere, $S_8$, of radius $8
h^{-1}$ Mpc are characterized by
\begin{equation}
\delta(\mathbf{x}) =
\frac{\rho(\mathbf{x})-
  \langle\rho\rangle_{S_8}}{\langle\rho\rangle_{S_8}},   
\label{delta}
\end{equation}
where $\langle \cdots \rangle_{S_8}$ denotes the spatial-average over
$S_8$. Of particular interest is $\sigma_8$, the rms fluctuations of
$\delta$ within $S_8$, 
\begin{equation}
\sigma_8^2 \equiv \langle (\delta(\mathbf{x})^2)\rangle_{S_8}
\label{sigma}
\end{equation}
Since $\sigma_8$ is used to study the formation of structure in the
universe, usual calculations of $\sigma_8$ involves modeling the
evolution of density fluctuations in the universe as it expands. 
These calculations are done in Fourier space,
and $\sigma_8$ is usually expressed in terms of a power spectrum,
$\Delta_k$, and a window function, $W(k)$ ($W(r)$ in real space), that
restricts the fluctuations in the density to the region $S_8$.  There
are a number of subtleties in this calculation, however, and in
particular, care must be taken in choosing $W(k)$. Indeed, the simplest
choice   
\[W(r)= \left\{ 
\begin{array}{l l}
  1 & \quad \mbox{for $r \le 8 h^{-1}$ Mpc}\\
  0 & \quad \mbox{for $r > 8 h^{-1}$ Mpc,}\\
\end{array} 
\right.
\] 
called the top hat window, produces a hard cutoff at $r = 8 h^{-1}$ Mpc,
and thus can introduce spurious fluctuations at scales $k \sim h/8$
Mpc$^{-1}$.  

Our focus is not on the formation of structure during the expansion of
the universe, however.  It is on using $\sigma_8$ to test the
predictions of the extended GEOM. As we have an explicit
density profile for a model galaxy, we can calculate 
$\sigma_8$ directly from its definition, Eq.~$(\ref{sigma})$, using the
top-hat window function; a power 
spectrum is not needed in our calculation.  

One estimate of the characteristic length scale
for the galaxy-galaxy correlation function \cite{TurnerBook} is $5
h^{-1}$; thus there is roughly one galaxy within $S_8$. Consequently,
we calculate $\sigma_8$ with the approximation that there is  
a single galaxy in $S_8$ with its core located at the center of the
sphere, and with a specific set of values, $v_H^{*}$ and $r_H^{*}$,
for the asymptotic velocity and core radius of this galaxy. 

As $\rho(r)$ varies significantly across $S_8$, we begin by finding the
average density,  
\begin{eqnarray}
\langle \rho\rangle_8 &\equiv& 
\langle \rho_H \theta(r_H - r)\rangle_8 +
\langle \rho_{\hbox{\scriptsize{asymp}}}(r)\theta(r-r_H)\rangle_8 + \langle
\rho_{II}^{1}(r)\theta(r-r_H)\rangle_8,
\nonumber
\\
&=&
3\left(\frac{1+\alpha_\Lambda}{1+3\alpha_\Lambda}\right)
\rho_{\hbox{\scriptsize{asymp}}}(u_8)D(u_8),
\label{ave-rho}
\end{eqnarray}
in $S_8$. Here, $\theta(x)$ is the step function, and ${u}_8 = 8
h^{-1}\hbox{Mpc}/\chi^{1/2}\lambda_{DE}$.  
The resultant average density is proportional to
$\rho_{\hbox{\scriptsize{asymp}}}(u_8)$, with 
\begin{eqnarray}
D(u_8) &\equiv & 1 - y_8^{\frac{1+3\alpha_\Lambda}{1+\alpha_\Lambda}} + 
\left(\frac{1+3\alpha_\Lambda}{1+\alpha_\Lambda}\right)\zeta
-
\frac{3}{2}(1+\alpha_{\Lambda})\left[\frac{1}{2}\widetilde{C}_{\cos} -
  \nu\widetilde{C}_{\sin}\right]y_8 + 
\nonumber
\\
&{}&
\frac{3}{2}(1+\alpha_\Lambda)y_8^{1/2}\Bigg(
\left[\frac{1}{2}\widetilde{C}_{\cos}-\nu\widetilde{C}_{\sin}\right]
\cos\left[\nu\log y_8\right] -  
\nonumber
\\
&{}&
\qquad\qquad\qquad\>\>
\left[\frac{1}{2}\widetilde{C}_{\sin} + \nu\widetilde{C}_{\cos}\right]
\sin\left[\nu\log y_8\right]
\Bigg). 
\end{eqnarray}
Here, $y_8=r_H/(8 h^{-1})$, and
\begin{equation}
\widetilde{C}_{\cos} \equiv \frac{2}{3}\zeta
-\frac{1}{3}y^{2\alpha_\Lambda/(1+\alpha_\Lambda)}_8, \qquad
\nu\widetilde{C}_{\sin} \equiv \frac{7}{3}\zeta
-\frac{1}{6}\left(\frac{1+5\alpha_\Lambda}{1+\alpha_\Lambda}\right)
y^{2\alpha_\Lambda/(1+\alpha_\Lambda)}_8.
\end{equation}
The parameter
\begin{equation}
\zeta=\frac{
2{u}_8^{-2\alpha_\Lambda/(1+\alpha_\Lambda)}
}{
\Sigma(\alpha_\Lambda)\chi(\alpha_\Lambda)
}
\left(\frac{v_H^{*}}{c}
\right)^2,
\end{equation}
depends explicitly on the detail structure of the galaxy. 
As $\zeta\sim  4\times 10^{-3}$ for $v_H^{*}=200$ km/s, $D(u_8) \approx
1$, and the asymptotic density dominates
$\langle\rho\rangle_8$, not the detail structure of the
galaxy.

This is not the case for $\langle\delta(\mathbf{x})^2\rangle_8$, which
involves the integration of $\rho^2(r)$ over $S_8$. Not
surprising, it is now the behavior of the density near the core that
is relevant. Indeed, we find 
\begin{eqnarray}
\sigma_8^2 =&{}& -1+
\frac{1}{\left[D(u_8)\right]^2} 
\left[\frac{(1+3\alpha_\Lambda)}{(1+\alpha_\Lambda)}\right]^2
\Bigg(\frac{\zeta^2}{3}\left[\frac{4}{y_8} - 1\right] +
\nonumber
\\
&{}&
\frac{1}{3}\left[\frac{\alpha_\Lambda+1}{3\alpha_\Lambda
  -1}\right]\left[1-y^{\frac{3\alpha_\Lambda-1}{\alpha_\Lambda+1}}_8\right]
+
\frac{2\zeta}{3}\left(\frac{\alpha_\Lambda + 1}{\alpha_\Lambda-1}\right)
\left(1-y_8^{\frac{\alpha_\Lambda-1}{\alpha_\Lambda+1}}\right)
-
\nonumber
\\
&{}&
\frac{4(\alpha_\Lambda-3)(\alpha_\Lambda+1)}{(\alpha_\Lambda - 3)^2 +
  4\nu^2(1+\alpha_\Lambda)^2} \Bigg\{\left[\widetilde{C}_{\cos} -
  \frac{2(1+\alpha_\Lambda)}{(\alpha_\Lambda-3)}\nu\widetilde{C}_{\sin}
  \right]y_8^{\frac{\alpha_\Lambda-1}{\alpha_\Lambda+1}} -
\nonumber
\\
&{}&
\qquad\qquad\qquad\qquad\qquad
y_8^{1/2}\Bigg(
\left[\widetilde{C}_{\cos} -
  \frac{2(1+\alpha_\Lambda)}{(\alpha_\Lambda-3)} \nu\widetilde{C}_{\sin}  
  \right]\cos[\nu\log y_8]-
\nonumber
\\
&{}&
\qquad\qquad\qquad\qquad\qquad\qquad
\left[\widetilde{C}_{\sin}+
  \frac{2(1+\alpha_\Lambda)}{(\alpha_\Lambda-3)}
  \nu\widetilde{C}_{\cos}\right]\sin[\nu\log y_8]\Bigg)\Bigg\}+ 
\nonumber
\\
&{}&
\frac{3\zeta}{y_8}\frac{1}{9/4+\nu^2}\Bigg\{
\left[\widetilde{C}_{cos}+\frac{2\nu}{3}\widetilde{C}_{sin}\right] -
y^{\frac{3}{2}}_8 \Bigg(\left[\widetilde{C}_{\cos} +
  \frac{2\nu}{3}\widetilde{C}_{\sin}\right]\cos[\nu\log y_8] +  
\nonumber
\\
&{}&
\qquad\qquad\qquad\qquad\qquad\qquad\qquad\quad
\left[\frac{2\nu}{3}\widetilde{C}_{\cos}-
  \widetilde{C}_{\sin}\right]\sin[\nu\log y_8]\Bigg)\Bigg\}+ 
\nonumber
\\
&{}&
\frac{3}{4}\left[\frac{1}{y_8}-y_8\right]\left[\widetilde{C}_{\cos}^2+
\widetilde{C}_{\sin}^2\right]+
\frac{3}{4(1+\nu^2)y_8}\left[\widetilde{C}_{\cos}^2-
\widetilde{C}_{\sin}^2+2\nu\widetilde{C}_{\cos}
\widetilde{C}_{\sin}\right]-  
\nonumber
\\
&{}&
\frac{3y_8}{4(1+\nu^2)}\Bigg\{
\left[\widetilde{C}_{\cos}^2-
\widetilde{C}_{\sin}^2+2\nu\widetilde{C}_{\cos}
\widetilde{C}_{\sin}\right]\cos[2\nu\log y_8]+
\nonumber
\\
&{}&
\qquad\qquad\quad\>
\left[\nu(\widetilde{C}_{\cos}^2-
\widetilde{C}_{\sin}^2)-2\widetilde{C}_{\cos}
\widetilde{C}_{\sin}\right]\sin[2\nu\log y_8]\Bigg\}\Bigg).
\label{sigma-8}
\end{eqnarray}

To obtain values for $v^{*}_H$ and $r^{*}_H$, we are guided by the
operational definition of 
$\sigma_8$ described in \cite{TurnerBook}. This definition involves 
choosing a $S_8$ centered at a given point in the sky, calculating the
average mass in it, shifting this center to another point on the sky,
repeating the measurement, and continuing until all points in the sky
is, in principle, covered. The set of all such averages then forms an
ensemble of such measurements, and the rms fluctuation in the mass can
then be calculated for this ensemble\footnote{
Measurements of the average mass in $S_8$ are necessarily done at
different times.  As such, a Markovian assumption must be made
that the measurements of the average mass over time is equivalent to
an ensemble of average masses made at equal times. This same
assumption must be made when determining the representative galaxy.}.
    
Consistent with this operational definition of $\sigma_8$, $v_{H}^{*}$
and $r_H^{*}$ should be for a representative spiral 
galaxy for the universe.  Such a galaxy would in
principle be found through a survey of spiral galaxies, which would
result in an ensemble of asymptotic rotation 
velocities and core radii for the galaxies in the universe. An average
for each parameter can then be taken, and identified with $v_{H}^{*}$ and
$r_H^{*}$, which in turn can be used to construct the representative
galaxy.  While such a survey has not yet been done, there exists in
the literature a large repository of measurements of galactic rotational
velocity curves and core radii \cite{Blok-1} - \cite{Math}. Taken as a
whole, these 1393 galaxies are reasonably random, and is likely
representative of the observed universe at large.  

Although there have been a many studies of galactic rotation curves in
the literature, what is needed here is both the rotational velocity
and the core radius of galaxies. This requires both a
measurement of the velocity as a function of the distance from the
center of the galaxy, and a fit of the data to some model of the
velocity curve. To our knowledge, this analysis has been done in four
places in the literature. (The study \cite{McGa2005} gives fits
to MOND rotation curves, but does not list values for $r_H$.) While
each of the data sets were obtained with similar physical techniques,
there are distinct differences in their selection of galaxies, in the
exact experimental techniques used, and in the models to which the
observed rotation curves are fitted. In fact, the Hubble
constant used by each is often different from one another, and from
the value of 73.2 km/s/Mpc given by WMAP. The reader is referred to
the specific papers for details on how these observations were
made. 

A number of models are in use in the literature to fit the observed
galactic rotation curves, and they all require at least two
parameters to model observations. Four different
models of these velocity curves are fitted to the data sets we use here,
and out of these, two of them can be idealized using
$v^{\hbox{\scriptsize{ideal}}}(r)$. For these two models there is a
one-to-one correspondence between the parameters used for the fit with
$v_H$ and $r_H$. Data from these fits are then averaged to determine $v_H^{*}$
and $r_H^{*}$. 

A summary of the data sets, and how $v_H$ and $r_H$ are obtained for
each are as follows:

\vskip 0.25in

\noindent \textit{de Blok et.~al.~Data Set:} 
  De Blok and coworkers made detailed measurements of 60 LSB galaxies
  \cite{McGa}, and fits of the pseudo-isothermal velocity curve were
  done for 30 of them \cite{Blok-1}. Later, another set of
  measurements of 26 LSB galaxies 
  were made by de Blok and Bosma \cite{Blok-2}, of which 24 are
  different from the 30 listed in \cite{Blok-1}. Both the data for the
  30 original galaxies, and the 24 subsequent galaxies are
  used here. Although the authors used various models
  for determining the mass-to-light ratio in their measurements, we
  will use the data that comes from the minimum disk model, as this was
  the one model used for all of the galaxies in this set.
  
  De Blok and coworkers were chiefly concern with modeling the
  density profiles of observed galaxies, and as such, a parameter for
  the profile, $\rho_H$, along with a parameter for the core radius,
  $R_C$, were used by them.  The asymptotic value for their rotational
  velocity is $\sqrt{4\pi G\rho_H R_C^2}$. Identifying this expression
  with $v_H$, and $\sqrt{3} R_C$ with $r_H$, we are 
  able to extract from the de Blok data sets values for $v_H$ and
  $r_H$ along with their standard error. The authors used a value
  of 75 km/s/Mpc for the Hubble constant.    
\vskip 12pt
\noindent\textit{CF Data Set:} In \cite{Cour}, Courteau presented
observations of the rotational velocity curves for over 300 northern
Sb-Sc UGC galaxies, and determined $r_H$ for each by fitting the
curves to three different models of the velocity, one of which,
$v^{\hbox{\scriptsize vcA}}(r)=(2v_C/\pi)\arctan\left(r/r_t\right)$,
is similar to the velocity curve for the pseudo-isothermal profile
used by de Blok and coworkers. Like the pseudo-isothermal curve,
$v^{\hbox{\scriptsize{vcA}}}(r)$ can be approximated by the idealized
velocity curve used here. In the limit $r\gg r_t$,
$v^{\hbox{\scriptsize vcA}}\approx v_C$, which sets $v_C = v_H$. In
the limit $r\ll r_t$, $v^{\hbox{\scriptsize vcA}}\approx v_C(2r/\pi
r_t)$, which sets $r_t = 2r_H/\pi$. 

A fit of these observations was also made to a velocity curve where
the steepness of the transition from the hub and the asymptotic
velocity curves could also be taken into account. The model curve
has a form in the $r\to0$ limit that not only disagrees with our
idealized profile in one specific case, this curve does not fit all the
galaxies analyzed by Courteau.  We therefore did not use
data from this fit.

Values for $v_C$ and $r_t$ for 351 galaxies was obtained through the
VizieR service (http://vizier.u-strasbg.fr/viz-bin/VizieR). The great
majority of the rotation velocity curves were based on single observations of
the galaxy; only 75 of these galaxies were measured multiple times,
with the majority of these galaxies being observed twice. The data set
reposited at VizieR contained these multiple measurements, and we have
averaged the value of $v_C$ and $r_t$ for the galaxy when multiple
measurements were done. The standard error in the
repeated measurements of a single galaxy can be extremely large; this
was recognized in \cite{Cour}. A value of 70 km/s/Mpc was used for the
Hubble constant by the author.  
\vskip 12pt

\noindent\textit{Mathewson et.~al.~Data Set:} In \cite{Math}, a survey
of the velocity curves of 1355 spiral galaxies in the southern sky was
reported. Later, the rotation velocity curves for these observations were
derived in \cite{Pers-1995} after folding, deprojecting, and
smoothing the Mathewson data. Each of these velocity curves are due to
a single observation. Courteau performed a fit of Mathewson's
observations to the $v^{vcA}$ curve for 958 of the
galaxies in \cite{Cour} using a Hubble constant of $70$ km/s/Mpc. The
results of Courteau's analysis is reposited in VizieR as well. 

Persic, Salucci, and Stel has proposed the Universal Rotation Curve (URC)
\cite{Pers-1995}, \cite{Pers-1996}, which has been used by them to analyze the
Mathewson data. In addition, this model was the third model used by Courteau in
\cite{Cour} to fit both his and Mathewson's data.  While the URC
asymptotically approaches a constant velocity, at small $r$ the URC has a $r^{0.66}$
behavior, which is different from the pseudo-isothermal curve,
the $v^{\hbox{\scriptsize{vcA}}}$ curve, and the idealized velocity curve
considered here. Although the URC has a different power-law
behavior at small $r$, the difference is small enough that it
is unknown how $\sigma_8$ will change if the URC is used in its
calculation instead of the $v^{\hbox{\scriptsize{ideal}}}$ curve used
here. We leave this for future research; for our current purposes,
we did not consider fits to this velocity curve.
\vskip 12pt

\textit{Rubin et.~al.~Data Set:}  In the early 1980s, Rubin and
  coworkers \cite{Rubin1980} - \cite{Rubin1985} presented a
  detailed study of the rotation curves of 16 Sa, 23 Sb, and 21 Sc
  galaxies. This was \textit{not} a random sampling of such galaxies. 
  Rather, these galaxies were deliberately chosen to span a
  specified range of Sa, Sb, and Sc galaxies, and as stated in
  \cite{Burs}, averaging values of the properties of the galaxies in
  this data set would have little meaning. These measurements can
  contribute to a combined data set of all four measurements, 
  however, and we have included them in our analysis. While values for the core
  radii were not given, measurements of the rotational velocity as a
  function of the distance to the center of the galaxy were; we are
  able to fit this data to the same pseudo-isothermal rotation curve used by de
  Blok, et.~al. Results of this fit is given in \textbf{Appendix
    A}. A Hubble constant of 50 km/s/Mpc was used by the authors. 

\vskip12 pt

Wanting to be as unbiased and as inclusive as possible, we have deliberately
\textit{not} culled through the data sets to select the cleanest of the
rotation curves. Nevertheless, we have had to remove the data for 27
galaxies from the data sets. A list of these galaxies and the reason
why they were removed are given in \textbf{Appendix B}, where we have
also listed any peculiarities of the four base data sets. 

While $v_H$ is easily identified for all four data sets, determining
$r_H$ is more complicated. For the 
de Blok et.~al.~data set, published values of $R_C$ was first scaled
by $75/73.2$ to account for differences in the Hubble constant; $r_H$
is then obtained using $r_H=\sqrt{3}R_C$. The same calculation was
made using the calculated values of $R_C$ from 
\textbf{Appendix A}, but with $50/73.2$ instead of $75/73.2$ to
account for differences in Hubble constants. For the CF and Mathewson
et.~al.~data sets, published values of $r_t$ are first scaled by
$70/73.2$ to account for differences in Hubble constants, and 
$r_H$ is then obtained through $r_H=\pi r_t/2$. 

The values of $v^{*}_H$ and $r^{*}_H$ are then calculated for three of the
four basic data sets.  As the Rubin et.~al.~data set was not
random, $v^{*}_H$ and $r^{*}_H$ was not separately calculated for this
data set. These measurements were instead included with data
from the other three sets to form a Combined
data set, and $v_H^{*}$ and $r_H^{*}$ were calculated for this 
set as well. The results of these calculations are giving in Table
\ref{summary}.  These values for $v^{*}_H$ and $r^{*}_H$ were then used to
calculate $\sigma_8$ using Eq.~$(\ref{sigma-8})$, and the results of
this calculation are given in this table as well. All the data sets
give a value for $\sigma_8$ that agree with the WMAP value at the 95\% confidence
level. Moreover, the values for $\sigma_8$ for all of these data
sets\textemdash calculated from four sets of observations taken over a
thirty-year span\textemdash agree with one another at the 95\%
confidence level as well. This is consistent with our expectation that the
sample of 1393 measurements of galactic velocity curves used here is
reasonably random and unbiased.  

\begin{table}
{\centering
\begin{tabular}{l|rr|r|rr}
\hline
\textit{Data Set} &   $v^{*}_H\>\quad$  &  $r^{*}_H\>\quad$ & $R_{200}\,\>\>$ & $\sigma_8\>\>\quad$ \\
\hline
deBlok et.~al. (53)           &  $119.0_{\pm 6.8}$ &   $3.62_{\pm 0.33}$ & $210_{\pm 110}$ & $0.65_{\pm 0.11}$ \\
CF (348)                      &  $179.1_{\pm 2.9}$ &   $7.43_{\pm 0.35}$ & $182_{\pm 84}$  & $0.92_{\pm 0.18}$ \\
Mathewson et.~al.~(935)       &  $169.5_{\pm 1.9}$ &  $15.19_{\pm 0.42}$ & $224_{\pm 50}$  & $0.654_{\pm 0.093}$ \\
Combined (1393)               &  $172.1_{\pm 1.6}$ &  $11.82_{\pm 0.30}$ & $206_{\pm 53}$  & $0.73_{\pm 0.12}$ \\
\hline
\end{tabular}
\par}
\caption{The $v_H^{*}$ (km/s) and $r_H^{*}$ (kpc) is listed for each
  data set along with the resultant $R_{200}$ (kpc) and $\sigma_8$.
  The number of data points in each data set is listed in parentheses.}   
\label{summary}
\end{table}  

With $v_H^{*}$ and $r_H^{*}$, it is possible to calculate
$R_{200}$ by taking in
Eq.~$(\ref{ave-rho})$ $u_8\to u_{200}=R_{200}/\chi^{1/2}\lambda_{DE}$
and $\langle\rho\rangle_8 = 200\rho_c$ where the critical density,
$\rho_c = 1.006^{-0.085}_{+0.088}\times 10^{-29}$ g/cm$^3$ from
\cite{WMAP}. The resultant equation is  
solved numerically for $R_{200}$; the results of this analysis is
given in Table $\ref{summary}$ as well. The range of values for
$R_{200}$\textemdash with $R_{200} =  206_{\pm 53}$  kpc for the
Combined data set\textemdash is consistent with observations. Notice
that like the values for $\sigma_8$, the values of $R_{200}$ for these
data sets also agree with one another at the 95\% confidence level.

For both $\sigma_8$ and $R_{200}$, over 90\% of the
error is due to the error in $\alpha_\Lambda$; this is true for all
the data sets. The values of $\sigma_8$ and $R_{200}$ are thus
sensitively dependent on $\alpha_\Lambda$. On the one hand, this
sensitivity is consistent with $\alpha_\Lambda$ being a power-law
exponent. On the other hand, the excellent agreement between our
calculated values of $\sigma_8$ and the WMAP value for $\sigma_8$ is
all the more compelling because of it. 
 
\section{Dynamics of Test Particles in the Model Galaxy}

\subsection{The Extended and Gravitational Potentials}

Note that Eq.~$(\ref{genEOM})$ may be written in terms of an effective
potential, $\mathfrak{V}(\mathbf{x})$, as $\ddot{\mathbf{x}} = -
\mathbf{\nabla}\mathfrak{V}$, where
\begin{equation} 
\mathfrak{V}(\mathbf{x}) = \Phi(\mathbf{x}) +
c^2\log\mathfrak{R}[4+8\pi\rho/\Lambda_{DE}].
\label{effPot}
\end{equation}
Because of the additional terms from the extension of the GEOM on the
right hand side of Eq.~$(\ref{genEOM})$, it is
\textit{not} the gravitational potential, $\Phi(\mathbf{x})$, that
determines the dynamics of massive particles; it is instead
$\mathfrak{V}(\mathbf{x})$. This is an important distinction.  Our
extension of the GEOM can drastically change the metric, and these
changes will have broad implications if it is $\Phi(\mathbf{x})$ and not 
$\mathfrak{V}(\mathbf{x})$ that determine the dynamics. To see this,
we calculate explicitly $\Phi(\mathbf{x})$ in 
Regions I and II. In Region III, $r>r_{II}$, and motion in this region
is not physically relevant.

Integrating $\mathbf{\nabla}^2\Phi = 4\pi G\rho$ in Region I gives
\begin{equation}
\Phi(r) = \Phi(0)+\frac{1}{2}v_H^2 \left(\frac{r}{r_H}\right)^2,
\end{equation}
where $\Phi(0)$ is an overall integration constant. That this is the usual
expression for the Newtonian potential can be seen from the relation
$\rho_H=3v_H^2/4\pi Gr_H^2$. For Region II, we find 
\begin{eqnarray}
\Phi(r) = &{}&
\Phi(0) + \frac{1}{2}v_H^2 + v_H^2\log\left[\frac{r}{r_h}\right] -
\nonumber
\\
&{}&
\frac{(1+\alpha_\Lambda)^2}{2(1+3\alpha_\Lambda)}
\left(1-\frac{r_H}{r}\right)
\left(c^2\chi\Sigma(\alpha_\Lambda)u_H^{\frac{2\alpha_\Lambda}{(1+\alpha_\Lambda)}}
- 6 v_H^2\right)+
\nonumber
\\
&{}& 
\frac{(1+\alpha_\Lambda)^2}{(1+3\alpha_\Lambda)}
\frac{c^2\chi}{4\alpha_\Lambda}\Bigg\{
\frac{r^2}{\chi\lambda_{DE}^2}\left(\frac{8\pi\rho_{\hbox{\scriptsize{asymp}}}(r)}{\Lambda_{DE}}-\alpha_\Lambda
  \frac{8\pi\rho_{II}^1(r)}{\Lambda_{DE}}\right) - 
\nonumber
\\
&{}&
\qquad\qquad\qquad
\frac{r^2_H}{\chi\lambda_{DE}^2}\left(\frac{8\pi\rho_{\hbox{\scriptsize{asymp}}}(r_H)}{\Lambda_{DE}}-\alpha_\Lambda
  \frac{8\pi\rho_{II}^1(r_H)}{\Lambda_{DE}}\right)\Bigg\}.
\label{potential}
\end{eqnarray}
The $1/r$ term in $\Phi(\mathbf{x})$ is expected from Newtonian
gravity, and is due to the boundary conditions for $\Phi$ at $r=r_H$;
this is true for the constant terms as well. The logarithmic term is
due specifically to the source, $f(r)$, as expected. It is a 
long-range potential that extends out to $r_{II}$, and could potentially
explain the non-Newtonian interaction observed between galaxies and galactic
clusters. The $\rho_{II}^{1}(r) r^2$ term is due to the perturbation
of the asymptotic density, and contains terms $\sim 1/r^{1/2}$. It is
due to both the boundary terms in $\rho_{II}^{1}$ and the boundary
conditions for $\Phi(r)$. 

For the $c^2$ term in $\Phi(r)$, $r^2\rho_{\hbox{\scriptsize{asymp}}} \sim
r^{2\alpha_\Lambda/(1+\alpha_\Lambda)}$, which increases as $r^{1.22}$
for $\alpha_\Lambda = 1.56$. This would dominate the dynamics of test
particles in the galaxy if 
the extended GEOM depended on $\Phi(\mathbf{x})$ instead of
$\mathfrak{V}(\mathbf{x})$. Indeed, if the dynamics were determined by
$\Phi(\mathbf{x})$, the resultant motion for stars in the galaxy would not
agree with observations whatsoever.  Instead, this term and the
$r^2\rho_{II}^1$ term in $\Phi$ are canceled by the additional
density-dependent terms in Eq.~$(\ref{effPot})$. To see this, in
Regions I and II $\rho >> \Lambda_{DE}/2\pi$, and we expand
Eq.~$(\ref{effPot})$ to give 
\begin{equation}
\mathfrak{V}(\mathbf{x}) = \Phi(0) + \frac{1}{2}v_H^2
\left(\frac{r}{r_H}\right)^2 + 
\frac{c^2\chi}{2\alpha_\Lambda}\left(\frac{\Lambda_{DE}}{8\pi\rho_H}\right)^{\alpha_\Lambda}, 
\end{equation}
in Region I, while in Region II,
\begin{eqnarray}
\mathfrak{V}(r) = &{}&
\Phi(0) + \frac{1}{2}v_H^2 + v_H^2\log\left[\frac{r}{r_H}\right] -
\nonumber
\\
&{}&
\frac{(1+\alpha_\Lambda)^2}{2(1+3\alpha_\Lambda)}
\left(1-\frac{r_H}{r}\right)
\left(c^2\chi\Sigma(\alpha_\Lambda)u_H^{\frac{2\alpha_\Lambda}{(1+\alpha_\Lambda)}}
- 6 v_H^2\right)-
\\
&{}&
\frac{(1+\alpha_\Lambda)^2}{(1+3\alpha_\Lambda)}
\frac{c^2\chi}{4\alpha_\Lambda}\Bigg\{ 
\frac{r^2_H}{\chi\lambda_{DE}^2}\left(\frac{8\pi\rho_{\hbox{\scriptsize{asymp}}}(r_H)}{\Lambda_{DE}}-\alpha_\Lambda
  \frac{8\pi\rho_{II}^1(r_H)}{\Lambda_{DE}}\right)\Bigg\},
\nonumber
\label{eff-pot}
\end{eqnarray}
where we have used $\rho_{II}^{1}/\rho_{\hbox{\scriptsize{asymp}}}<<1$ and
Eq.~$(\ref{amp})$. The effective potential thus increases only
logarithmically as $r$ increases. This is expected, and is consistent
with using the rotational velocity curves in constructing $f(r)$. 

The $r^{1.22}$ increase in $\Phi(r)$, if unchecked, would mean that
at some point the weak gravity approximation used here would be
violated. Indeed, it is doubtful if the resultant spacetime will be
gravitationally stable, and it would certainly be inconsistent with WMAP
measurements, which indicate that the universe is flat. We found that
Region III the density $\rho\to 0$ exponentially fast, however, and thus
$\Phi\to 0$ exponentially fast as well when $r>r_{II}$.  This provides
a cutoff to $\Phi(r)$, and prevents the gravitational potential from
becoming infinite at large $r$.  The growth in $\Phi$ ends at
$r=r_{II}$, and as we find that $\Phi(r_{II})-\Phi(0)\approx 0.02
c^2$, the weak gravity approximation is valid for Regions I and II.

\subsection{Matter Density Measurements Under the Extended GEOM}

That the motion of stars in the galaxy is determined by
$\mathfrak{V}(\mathbf{x})$ and not by $\Phi(\mathbf{x})$ has far
reaching implications. The local density of matter in the universe is
determine through observations on the influence that this density has
on the motion of test particles. These test particles can be divided
into two classes:  massive particles, such as the motion of stars in
galaxies, and massless particles, such as the motion of photons of
various frequencies. 

The extended GEOM affects only the motion of massive test particles.
As noted above, the motion of such particles is determined by the
effective potential, $\mathfrak{V}$.  As $\mathfrak{V}$ is in turn
determined by $\rho_{II}^1$, and not the background density,
$\rho_{\hbox{\scriptsize{asymp}}}$, observations of the
rotational velocity curves of a galaxy using the motion of the stars
in the galaxy will at best be able to determine the perturbation on the
background density, $\rho_{II}^1$, and not
$\rho_{\hbox{\scriptsize{asymp}}}$ itself. Since
$\rho_{\hbox{\scriptsize{asymp}}}(r) >> \rho_{II}^1(r)$ when $r>>r_H$,
\emph{the majority of the mass in the universe cannot be seen with
  these methods}. In particular, the motion of stars in galaxies can
only be used to estimate $\rho(r)-\rho_{\hbox{\scriptsize{asymp}}}$;
the matter in $\rho_{\hbox{\scriptsize{asymp}}}$ is present, but
cannot be determined in this way.  

The extended GEOM does not affect the motion of massless particles,
however. Thus, the trajectory of photons still follow the GEOM, and
these equations are determined by the local 
metric, $g_{\mu\nu}$. In the nonrelativistic and weak gravity limits,
$g_{\mu\nu}$ is determined in turned by the potential, $\Phi(\mathbf{x})$.
Consequently, measurements of the local density of matter using
photons\textemdash such as through gravitational lensing\textemdash
results in determining the total density, $\rho$, and not just the small
fraction of it in $\rho_{II}^1$ as is possible using massive
particles. 

\section{Concluding Remarks}

Given how sensitive our expression for $\sigma_8$ is dependent on
$\alpha_\Lambda$, that our predicted values of 
$\sigma_8$ is within experimental error of its measured value by WMAP is a
compelling result. This is especially true as the data used in calculating
$\sigma_8$ was taken by four different groups over 
a period of thirty years, and for purposes that have no connection
whatsoever with our analysis. In the absence of a direct
experimental measurement of $\alpha_\Lambda$, this agreement between
the calculated and measured values of $\sigma_8$ provides a
persuasive argument for the validity of our extension of the GEOM. 

In \cite{ADS}, we performed a gedanken experiment based on the
measurement of anomalous accelerations in a dilute gas due to the
passage of a sound wave with wavenumber $k=1$ cm$^{-1}$, and an
amplitude that is 10\% that of an ambient gas that has a density of
$10^{-18}$ g/cm$^3$. We were able to establish a rough lower bound on
$\alpha_{\Lambda}$ that for 
$\Lambda_{DE} = 7.21 \times 10^{-30}$ g/cm$^3$ is
$1.56$. This is equal to the value of $\alpha_\Lambda$ found in
Sec.~3.  While this gedanken 
experiment was a crude estimate of the effects of the extended GEOM
and while it is an open question as to whether such an experiment
would be feasible, the fact that the value of
$\alpha_\Lambda$ found is so close to this lower bound raises the
possibility that direct detection and measurement of
$\alpha_\Lambda$ through terrestrial experiments may be possible in
the near future. 

\appendix{}

\section{Fitting the Rubin et.~al.~Data Set}

In \cite{Rubin1985}, measurements of the rotational velocity as a
function of radius for 60 Sa, Sb and Sc spiral galaxies are given.
These measurements allows us to fit the data given to a model of the
rotational velocity curve. Instead of fitting the data directly to the
pseudo-isothermal velocity curve,
$v^{\hbox{\scriptsize p-iso}}(r)$, as is done in \cite{Blok-1}, it is
more convenient to fit it to $(v^{\hbox{\scriptsize
    p-iso}}(r))^2$. Moreover, since what is needed is the asymptotic 
rotational velocity instead of the density parameter for the
pseudo-isothermal profile, our fit to $(v^{\hbox{\scriptsize
    p-iso}}(r))^2 = v_H^2 c(r)$ uses $v_H$ and $R_C$ as the
two parameters. Here, 
\begin{equation}
c(r) = 1-\frac{R_C}{r} \arctan\left(\frac{r}{R_C}\right).
\end{equation}
The variance of the fit is then
\begin{equation}
\sigma_{(v^{\hbox{\scriptsize p-iso}})^2}^2 \equiv
\frac{1}{N-2}\sum_{n=1}^N [(v^{\hbox{\scriptsize p-iso}}_n)^2 - v_H^2c(r_n)]^2,
\label{D}
\end{equation}
where $\{(v_n^{\hbox{\scriptsize{p-iso}}}, r_n)\}$ is the set $N$ of rotational
velocity verses radius measurements for a galaxy. A least squares fit gives
\begin{equation}
v_H^2 = \frac{\langle (v^{\hbox{\scriptsize p-iso}}_n)^2 c(r_n)\rangle}{\langle
  c(r_n)^2\rangle},
\label{v-infty}
\end{equation}
where $\langle \cdots\rangle$ denotes an average over the data
points. While an equation for $R_C$ can also be given through minimization
of Eq.~$(\ref{D})$, the resultant equation for $R_C$ is given
implicitly.  We find it more useful to substitute
Eq.~$(\ref{v-infty})$ into Eq.~$(\ref{D})$, and then use iteration to
find the $R_C$ that minimizes $\sigma^2_{(v^{\hbox{\scriptsize
      p-iso}})^2}$. Using this value for $R_C$, $v_H^2$ is then given 
by Eq.~$(\ref{v-infty})$. 

The standard error, $\sigma_{v_H}$, in $v_H$, is
\begin{eqnarray}
\sigma_{v_H}= &{}& \frac{1}{2}\Bigg(\frac{\sigma^2_{(v^{\hbox{\scriptsize p-iso}}_n)^2}}{
  Nv_H^2\langle c(r_n)^2\rangle} + 
\nonumber
\\
&{}&
\qquad
\left[1 + \frac{1}{v_H^2\langle c(r_n)^2\rangle}
\left\langle\frac{v^{\hbox{\scriptsize{p-iso}}}_nr_n^2}{R_C^2+r_n^2}\right\rangle 
-
\frac{2}{\langle c(r_n)^2\rangle}\left\langle\frac{c(r_n)r_n^2}{R_c^2+r_n^2}\right\rangle
\right]^2\frac{\sigma_{R_C}^2}{R_c^2}
\Bigg)^{\frac{1}{2}},
\end{eqnarray}
while the standard error, $\sigma_{R_C}$, in $R_C$ is
\begin{equation}
\sigma_{R_C}= \frac{R_C\> \sigma_{(v^{\hbox{\scriptsize p-iso}}_n)^2}\sqrt{\langle
    c(r_n)^2\rangle}}{\Delta\sqrt{N}} \left\{\langle
c(r_n)^2\rangle\left\langle\frac{r_n^4}{(r_n^2+R_C^2)^2}\right\rangle 
-\left\langle\frac{c(r_n)r_n^2}{r_n^2+R_C^2}\right\rangle^2 \right\}^{1/2}.
\end{equation}
Here, 
\begin{eqnarray}
\Delta \equiv &{}& 
2\langle (v^{\hbox{\scriptsize p-iso}}_n)^2c(r_n)\rangle
\left\langle\frac{c(r_n)r_n^4}{(r_n^2+R_C^2)^2}\right\rangle -
2\langle c(r_n)^2\rangle
\left\langle\frac{(v^{\hbox{\scriptsize p-iso}}_n)^2
  r_n^4}{(r_n^2+R_C^2)^2}\right\rangle + 
\nonumber
\\
&{}&
\left\langle \frac{c(r_n)r_n^2}{r_n^2+R_C^2}\right\rangle
\left\langle\frac{(v^{\hbox{\scriptsize p-iso}}_n)^2
  r_n^2}{r_n^2+R_C^2}\right\rangle - 
\langle (v^{\hbox{\scriptsize p-iso}}_n)^2 c(r_n)\rangle
\left\langle\frac{r_n^4}{(r_n^2+R_C^2)^2}\right\rangle.
\end{eqnarray}

The results of our fits of the Rubin et.~al.~data are tabulated in Table
\ref{Rubin}. The base data from \cite{Rubin1985} used
a Hubble constant of $50$ km/s/Mpc, and the results given in the table
are for this value of the constant. Of the 60
galaxies from \cite{Rubin1985}, NGC 6314 and IC 724 could 
not be fitted to a nonzero $R_C$, while the fit for NGC  2608 resulted
in a $R_C$ that is less than $0.01$ kpc. 

\begin{table}
{
\centering
\begin{tabular}{l|rrrr||l|rrrr}
\hline
Galaxy &  $R_C$ & \hskip-0.2in $\>\>\Delta R_C$ & \hskip-0.2in $v_H\>\>$ & \hskip-0.2in $\Delta v_H$ & Galaxy  &
$R_C$ &  \hskip-0.2in $\>\>\Delta R_C$ & \hskip-0.2in $v_H\>\>$ & \hskip-0.2 in$\Delta v_H$   \\
\hline
NGC 1024 &   0.27 &   0.14 &   229.42 &    9.77 &  NGC 4800 &   0.18 &   0.06 &   171.56 &   3.42    \\
NGC 1357 &   0.52 &   0.14 &   268.19 &   16.27 &  NGC 7083 &   0.89 &   0.14 &   226.51 &   2.27    \\
NGC 2639 &   1.02 &   0.34 &   337.69 &   31.31 &  NGC 7171 &   2.25 &   0.36 &   251.35 &   6.47    \\
NGC 2775 &   0.40 &   0.17 &   298.98 &    8.90 &  NGC 7217 &   0.19 &   0.10 &   275.21 &   6.56    \\
NGC 2844 &   0.41 &   0.09 &   167.50 &   18.93 &  NGC 7537 &   0.80 &   0.10 &   150.06 &   2.35    \\
NGC 3281 &   0.44 &   0.05 &   211.32 &   26.51 &  NGC 7606 &   1.40 &   0.30 &   279.29 &   4.11    \\
NGC 3593 &   0.16 &   0.08 &   115.28 &   14.01 &  UGC 11810&   1.54 &   0.38 &   193.28 &   3.85    \\
NGC 3898 &   0.53 &   0.06 &   254.76 &   28.73 &  UGC 12810&   3.22 &   0.35 &   245.73 &   1.47    \\
NGC 4378 &   0.13 &   0.06 &   307.61 &   26.60 &  NGC 701  &   2.49 &   0.58 &   188.78 &   4.86    \\
NGC 4419 &   0.63 &   0.03 &   211.55 &    2.33 &  NGC 753  &   0.31 &   0.11 &   208.50 &   3.57    \\
NGC 4594 &   1.65 &   0.30 &   397.24 &   10.15 &  NGC 801  &   0.79 &   0.16 &   227.64 &   4.06    \\
NGC 4698 &   1.85 &   0.47 &   284.96 &    6.34 &  NGC 1035 &   1.24 &   0.09 &   150.62 &   1.26    \\
NGC 4845 &   0.11 &   0.07 &   187.54 &    0.07 &  NGC 1087 &   0.54 &   0.10 &   131.91 &   2.54    \\
UGC 10205&   2.19 &   0.27 &   272.34 &    4.07 &  NGC 1421 &   0.54 &   0.13 &   176.42 &   3.94    \\
NGC 1085 &   0.29 &   0.05 &   307.02 &    2.11 &  NGC 2715 &   1.10 &   0.22 &   151.47 &   2.93    \\
NGC 1325 &   1.80 &   0.28 &   195.55 &    2.67 &  NGC 2742 &   1.10 &   0.16 &   181.86 &   2.36    \\
NGC 1353 &   0.36 &   0.18 &   218.48 &    8.30 &  NGC 2998 &   1.08 &   0.22 &   213.85 &   3.22    \\
NGC 1417 &   0.40 &   0.05 &   278.87 &    2.36 &  NGC 3495 &   3.11 &   0.46 &   206.75 &   3.22    \\
NGC 1515 &   0.06 &   0.10 &   178.35 &   10.03 &  NGC 3672 &   1.74 &   0.24 &   208.11 &   4.03    \\
NGC 1620 &   1.73 &   0.25 &   241.62 &    3.14 &  NGC 4062 &   0.79 &   0.13 &   167.88 &   2.65    \\
NGC 2590 &   1.30 &   0.54 &   255.24 &    5.33 &  NGC 4321 &   0.79 &   0.35 &   208.24 &   5.42    \\
NGC 2708 &   1.91 &   0.68 &   269.92 &    9.45 &  NGC 4605 &   0.97 &   0.32 &   112.62 &   3.42    \\
NGC 2815 &   1.91 &   0.68 &   269.92 &    9.45 &  NGC 4682 &   1.17 &   0.23 &   181.17 &   2.97    \\
NGC 3054 &   2.41 &   0.56 &   259.10 &    8.30 &  NGC 7541 &   0.21 &   0.16 &   195.04 &   5.94    \\
NGC 3067 &   0.76 &   0.06 &   156.80 &    1.22 &  NGC 7664 &   0.65 &   0.14 &   196.05 &   3.07    \\
NGC 3145 &   0.15 &   0.07 &   257.00 &    4.84 &  IC 467   &   1.64 &   0.33 &   152.42 &   3.26    \\
NGC 3200 &   0.42 &   0.09 &   266.07 &    5.43 &  UGC 2885 &   0.06 &   0.10 &   266.22 &   5.88    \\
NGC 3223 &   1.35 &   0.23 &   275.29 &    5.51 &  UGC 3691 &   3.04 &   0.33 &   229.42 &   1.31    \\
NGC 4448 &   0.59 &   0.11 &   207.02 &    1.98 & & & & & {} \\
\hline
\end{tabular}
\par
}
\caption{Fitted values of $R_C$ (kpc) and $v_H$ (km/s), and their errors for
  the Rubin et.~al.~data set.}  
\label{Rubin}
\end{table}

\section{Data sets}

For the de Blok et.~al.~data set, the galaxy F568-3 was analyzed
twice; we use the analysis of F568-3 given by the authors
in \cite{Blok-1}. In de Blok and Bosma \cite{Blok-2}, two of the
galaxies, F563-1 and U5750, also appeared in \cite{Blok-1}; we used
the values from \cite{Blok-1} for these galaxies in our analysis.
Finally, the radius for DDO185 from \cite{Blok-1} was not determined,
and we could not include this data point in our analysis. Thus, out of
56 galaxies in this data set, 53 were used.   

For the Rubin et.~al.~data set, we could not find a nonzero radius for
two galaxies, and one galaxy had a radius less than 0.01 kpc. As this
radius was smaller than the resolution of their observations, this
data point was not included. A total of 57 galaxies were thus used from
\cite{Rubin1985}.  

For the CF and Mathewson et.~al.~data sets, the vast majority of the
data were based on single observations. While we may have greater
leeway in removing outliers from this data set, even here we were
circumspect. First, 75 galaxies in \cite{Cour} were observed multiple
times. Of these, the galaxies UGC 7234 and UGC 10096 had listed an
asymptotic velocity for one of the observations that was opposite from
the measured asymptotic velocity for the others. We assumed that this
was a typographical error, and the sign of the anomalous rotational
velocity is reversed. Second, five galaxies in 
the CF and Mathewson et.~al.~data sets had a $r_H=0$, one galaxy had a
radius that was 11-sigma out from the mean, and three galaxies had a $v_H$
that exceeded 8,000 km/s. These are likely indications that the data
was not sufficiently accurate to allow for a fit of the velocity
curve, and these galaxies were removed from the data sets. Finally,
given that there are only 1393 galaxies combined in the data sets, if
a galaxy had a $v_H$ or a $r_H$ that was five-sigma or more out from
the mean, it was removed. In the end, 348 galaxies were used in the CF
data set, and 935 galaxies were used in the Mathewson et.~al.~data
set. A summary of the data points not used in our analysis is given in
Table \ref{removed}.  

\begin{table}
{\centering
\begin{tabular}{l|l|l}
\hline
\textit{Data Set} & \textit{Data Removed} & \textit{Reason} \\
\hline
de Blok et.~al.   & DDO185      &   $r_H=\infty$            \\                  
Rubin et.~al.     & NGC 6314    &   $r_H = 0$               \\                
{}                & IC 724      &   $r_H = 0$               \\                
{}                & NGC 2608    &   $r_H < 0.01$            \\                
CF                & UGC 6534    &   $v_H$ is 35 $\sigma$ out\\                
{}                & UGC 12543   &   $v_H$ is 11 $\sigma$ out\\                
Mathewson et.~al.~& ESO 140-G28 &   $v_H > 8,000$ km/s      \\        
{}                & ESO 481-G30 &   $v_H > 24,000$ km/s     \\                
{}                & ESO 443-G42 &   $v_H > 94,000$ km/s     \\               
{}                & ESO 108-G19 &   $r_H = 0$               \\                
{}                & ESO 141-G34 &   $r_H = 0$               \\               
{}                & ESO 21-G5   &   $r_H$ is 6 $\sigma$ out \\            
{}                & ESO 548-G21 &   $r_H$ is 7 $\sigma$ out \\              
{}                & NGC 7591    &   $r_H = 0$               \\               
{}                & ESO 243-G34 &   $v_H$ is 5 $\sigma$ out \\ 
{}                & ESO 317-G41 &   $r_H = 0$               \\  
{}                & ESO 358-G9  &   $v_H$ is 6 $\sigma$ out \\  
{}                & ESO 435-G25 &   $v_H$ is 5 $\sigma$ out \\   
{}                & ESO 467-G12 &   $r_H = 0$               \\  
{}                & ESO 554-G28 &   $v_H$ is 6 $\sigma$ out \\  
{}                & ESO 60-G24  &   $v_H$ is 10 $\sigma$ out\\
{}                & ESO 359-G6  &   $r_H$ is 11 $\sigma$ out\\
{}                & ESO 481-G21 &   $r_H$ is 6 $\sigma$ out \\
{}                & UGCA 394    &   $r_H$ is 7 $\sigma$ out \\
{}                & ESO 298-G15 &   $ r_H$ is 7 $\sigma$ out\\
{}                & ESO 545-G3  &   $r_H$ is 7 $\sigma$ out \\
{}                & ESO 404-G18 &   $r_H$ is 9 $\sigma$ out \\
\hline
\end{tabular}
\par}
\centering
\caption{Listed are the galaxies removed from the data sets used in
  our analysis along with the reason for their removal.}
\label{removed}
\end{table}


\begin{thebibliography}{21}
\expandafter\ifx\csname natexlab\endcsname\relax\def\natexlab#1{#1}\fi
\expandafter\ifx\csname bibnamefont\endcsname\relax
    \def\bibnamefont#1{#1}\fi
\expandafter\ifx\csname bibfnamefont\endcsname\relax
    \def\bibfnamefont#1{#1}\fi
\expandafter\ifx\csname citenamefont\endcsname\relax
    \def\citenamefont#1{#1}\fi
\expandafter\ifx\csname url\endcsname\relax
    \def\url#1{\texttt{#1}}\fi
\expandafter\ifx\csname urlprefix\endcsname\relax\def\urlprefix{URL }\fi
\providecommand{\bibinfo}[2]{#2}
\providecommand{\eprint}[2][]{\url{#2}}

\bibitem[1]{ADS}
\bibinfo{author}{\bibfnamefont{A.~D.}~\bibnamefont{Speliotopoulos}},
{\bibinfo{journal}{Gen.~Relativ.~Gravit.}
    \textbf{\bibinfo{volume}{42}}, \bibinfo{pages}{1537} (\bibinfo{year}{2010})}.

\bibitem[2]{Ries1998}
\bibinfo{author}{\bibfnamefont{A.~G.}~\bibnamefont{Riess}},
        {\bibfnamefont{A.~V.}~\bibfnamefont{Filippenko}},
        {\bibfnamefont{P.}~\bibnamefont{Challis}},
        {\bibfnamefont{A.}~\bibfnamefont{Clocchiatti}},
        {\bibfnamefont{A.}~\bibfnamefont{Diercks}},
        {\bibfnamefont{P.~M.}~\bibfnamefont{Garnavich}},
          {\bibfnamefont{R.~L..}~\bibfnamefont{Gilliland}},
            {\bibfnamefont{C.~J.}~\bibfnamefont{Hogan}},
            {\bibfnamefont{S.}~\bibfnamefont{Jha}},
              {\bibfnamefont{R.~P.}~\bibfnamefont{Kirshner}},
                {\bibfnamefont{B.}~\bibfnamefont{Leibundgut}},
                  {\bibfnamefont{M.~M.}~\bibfnamefont{Phillips}},
                    {\bibfnamefont{D.}~\bibfnamefont{Riess}},
                      {\bibfnamefont{B.~P.}~\bibfnamefont{Schmidt}},
                      {\bibfnamefont{R.~A.}~\bibfnamefont{Schommer}},
                      {\bibfnamefont{R.~C.}~\bibfnamefont{Smith}},
                    {\bibfnamefont{J.}~\bibfnamefont{Spyromilio}},
                      {\bibfnamefont{C.}~\bibfnamefont{Stubbs}},
                      {\bibfnamefont{N.~B.}~\bibfnamefont{Suntzeff}},
                      \bibnamefont{and}
                      {\bibfnamefont{J.}~\bibfnamefont{Tonry}}, 
    \bibinfo{journal}{Astron.~J.} \textbf{\bibinfo{volume}{116}},
    \bibinfo{pages}{1009} (\bibinfo{year}{1998}).

\bibitem[3]{Perl1999}
\bibinfo{author}{\bibfnamefont{S.}~\bibnamefont{Perlmutter}},
        {\bibfnamefont{G.}~\bibnamefont{Aldering}},
        {\bibfnamefont{G.}~\bibfnamefont{Goldhaber}},
        {\bibfnamefont{R.~A.}~\bibfnamefont{Knop}},
        {\bibfnamefont{P.}~\bibfnamefont{Nugent}},
          {\bibfnamefont{P.~G.}~\bibfnamefont{Castro}},
            {\bibfnamefont{S.}~\bibfnamefont{Deustua}},
            {\bibfnamefont{S.}~\bibfnamefont{Fabbro}},
              {\bibfnamefont{A.}~\bibfnamefont{Goobar}},
                {\bibfnamefont{D.~E.}~\bibfnamefont{Groom}},
                  {\bibfnamefont{I.~M.}~\bibfnamefont{Hook}},
                    {\bibfnamefont{A.~G.}~\bibfnamefont{Kim}},
                      {\bibfnamefont{M.~Y..}~\bibfnamefont{Kim}},
                      {\bibfnamefont{J.~C.}~\bibfnamefont{Lee}},
                      {\bibfnamefont{N.~J.}~\bibfnamefont{Nunes}},
                    {\bibfnamefont{R.}~\bibfnamefont{Pain}},
                      {\bibfnamefont{C.~R.}~\bibfnamefont{Pennypacker}},
                      {\bibfnamefont{R.}~\bibfnamefont{Quimby}},
                        {\bibfnamefont{C.}~\bibfnamefont{Lidman}},
                        {\bibfnamefont{R.}~\bibfnamefont{Ellis}},
                        {\bibfnamefont{M.}~\bibfnamefont{Irwin}},
                        {\bibfnamefont{R.~G}~\bibfnamefont{McMahon}},
                      {\bibfnamefont{P.}~\bibfnamefont{Ruiz-Lapuente}},
                      {\bibfnamefont{N.}~\bibfnamefont{Walton}},
                        {\bibfnamefont{B.}~\bibfnamefont{Schaefer}},
                        {\bibfnamefont{B.~J.}~\bibfnamefont{Boyle}},
                        {\bibfnamefont{A.~V.}~\bibfnamefont{Filippenko}},
                        {\bibfnamefont{P.}~\bibfnamefont{Matheson}},
                        {\bibfnamefont{A.~S.}~\bibfnamefont{Fruchter}},
                        {\bibfnamefont{N.}~\bibfnamefont{Panagia}},
                        {\bibfnamefont{H.~J.~M.}~\bibfnamefont{Newberg}},
                        \bibnamefont{and}
                                    {\bibfnamefont{W.~J.}~\bibfnamefont{Couch}}, 
    \bibinfo{journal}{Astrophys.~J.~Suppl.} \textbf{\bibinfo{volume}{517}},
    \bibinfo{pages}{565} (\bibinfo{year}{1999}).

\bibitem[4]{WMAP}
\bibinfo{author}{\bibfnamefont{D.~N.}~\bibnamefont{Spergel}},
        {\bibfnamefont{R.}~\bibnamefont{Bean}},
        {\bibfnamefont{O.}~\bibfnamefont{Dor\'e}},
        {\bibfnamefont{M.~R.}~\bibfnamefont{Nolta}},
        {\bibfnamefont{C.~L.}~\bibfnamefont{Bennett}},
          {\bibfnamefont{J.}~\bibfnamefont{Dunkley}},
            {\bibfnamefont{G.}~\bibfnamefont{Hinshaw}},
            {\bibfnamefont{N.}~\bibfnamefont{Jarosik}},
              {\bibfnamefont{E.}~\bibfnamefont{Komatsu}},
                {\bibfnamefont{L.}~\bibfnamefont{Page}},
                  {\bibfnamefont{H.~V.}~\bibfnamefont{Peiris}},
                    {\bibfnamefont{L.}~\bibfnamefont{Verde}},
                      {\bibfnamefont{M.}~\bibfnamefont{Halpern}},
                      {\bibfnamefont{R.~S.}~\bibfnamefont{Hill}},
                      {\bibfnamefont{A.}~\bibfnamefont{Kogut}},
                    {\bibfnamefont{M.}~\bibfnamefont{Limon}},
                      {\bibfnamefont{S.~S.}~\bibfnamefont{Meyer}},
                        {\bibfnamefont{N.}~\bibfnamefont{Odegard}},
                        {\bibfnamefont{G.~S.}~\bibfnamefont{Tucker}},
                        {\bibfnamefont{J.~L.}~\bibfnamefont{Weiland}},
                        {\bibfnamefont{E.}~\bibfnamefont{Wollack}},
                        \bibnamefont{and}
                                    {\bibfnamefont{E.~L.}~\bibfnamefont{Wright}}, 
    \bibinfo{journal}{Astrophys.~J.~Suppl.} \textbf{\bibinfo{volume}{170}},
    \bibinfo{pages}{277} (\bibinfo{year}{2007}).

\bibitem[5]{Blok-1}
\bibinfo{author}
{\bibfnamefont{W.~J.~G.}~\bibnamefont{de Blok}}, 
{\bibfnamefont{S.~S.}~\bibnamefont{McGaugh}},
{\bibfnamefont{A.}~\bibnamefont{Bosma}},
\bibnamefont{and}
{\bibfnamefont{V.~C.}~\bibnamefont{Rubin}},
    \bibinfo{journal}{Astrophys.~J.} \textbf{\bibinfo{volume}{552}},
    \bibinfo{pages}{L23} (\bibinfo{year}{2001}).

\bibitem[6]{Blok-2}
\bibinfo{author}
{\bibfnamefont{W.~J.~G.}~\bibnamefont{de Blok}},
\bibnamefont{and}
{\bibfnamefont{A.}~\bibnamefont{Bosma}},
    \bibinfo{journal}{Astro. Astrophys.} \textbf{\bibinfo{volume}{385}},
    \bibinfo{pages}{816} (\bibinfo{year}{2002}).

\bibitem[7]{McGa}
\bibinfo{author}
{\bibfnamefont{S.~S.}~\bibnamefont{McGaugh}},
{\bibfnamefont{V.~C.}~\bibnamefont{Rubin}},
\bibnamefont{and}
{\bibfnamefont{W.~J.~G.}~\bibnamefont{de Blok}}, 
    \bibinfo{journal}{Astron.~J.} \textbf{\bibinfo{volume}{122}},
    \bibinfo{pages}{2381} (\bibinfo{year}{2001}).

\bibitem[8]{Rubin1980}
\bibinfo{author}
{\bibfnamefont{V.~C.}~\bibnamefont{Rubin}}, 
{\bibfnamefont{W.~K.}~\bibnamefont{Ford, Jr.}},
\bibnamefont{and}
{\bibfnamefont{N.}~\bibnamefont{Thonnard}},
    \bibinfo{journal}{Astrophys.~J.} \textbf{\bibinfo{volume}{238}},
    \bibinfo{pages}{471} (\bibinfo{year}{1980}).

\bibitem[9]{Rubin1982}
\bibinfo{author}
{\bibfnamefont{V.~C.}~\bibnamefont{Rubin}}, 
{\bibfnamefont{W.~K.}~\bibnamefont{Ford, Jr.}},
{\bibfnamefont{N.}~\bibnamefont{Thonnard}},
\bibnamefont{and}
{\bibfnamefont{D.}~\bibnamefont{Burstein}},
    \bibinfo{journal}{Astrophys.~J.} \textbf{\bibinfo{volume}{261}},
    \bibinfo{pages}{439} (\bibinfo{year}{1982}).

\bibitem[10]{Burs}
\bibinfo{author}
{\bibfnamefont{D.}~\bibnamefont{Burstein}},
{\bibfnamefont{V.~C.}~\bibnamefont{Rubin}}, 
{\bibfnamefont{N.}~\bibnamefont{Thonnard}},
\bibnamefont{and}
{\bibfnamefont{W.~K.}~\bibnamefont{Ford, Jr.}},
    \bibinfo{journal}{Astrophys.~J., part 1} \textbf{\bibinfo{volume}{253}},
    \bibinfo{pages}{70} (\bibinfo{year}{1982}).

\bibitem[11]{Rubin1985}
\bibinfo{author}
{\bibfnamefont{V.~C.}~\bibnamefont{Rubin}}, 
{\bibfnamefont{D.}~\bibnamefont{Burstein}},
{\bibfnamefont{W.~K.}~\bibnamefont{Ford, Jr.}},
\bibnamefont{and}
{\bibfnamefont{N.}~\bibnamefont{Thonnard}},
    \bibinfo{journal}{Astrophys.~J.} \textbf{\bibinfo{volume}{289}},
    \bibinfo{pages}{81} (\bibinfo{year}{1985}).

\bibitem[12]{Cour}
\bibinfo{author}
{\bibfnamefont{S.}~\bibnamefont{Courteau}},
    \bibinfo{journal}{Astron.~J.} \textbf{\bibinfo{volume}{114}},
    \bibinfo{pages}{2402} (\bibinfo{year}{1997}).

\bibitem[13]{Math}
\bibinfo{author}
{\bibfnamefont{D.~S}~\bibnamefont{Mathewson}},
{\bibfnamefont{V.~L.}~\bibnamefont{Ford}},
\bibnamefont{and}
{\bibfnamefont{M.}~\bibnamefont{Buchhorn}},
    \bibinfo{journal}{Astrophys.~J.~Suppl.} \textbf{\bibinfo{volume}{82}},
    \bibinfo{pages}{413} (\bibinfo{year}{1992}).

\bibitem[14]{PeeblesBook}
\bibinfo{author}{\bibfnamefont{P.~J.~E.}
\bibnamefont{Peebles}} \bibnamefont{and}
\emph{\bibinfo{title}{The Large-Scale Structure of the Universe}}, Chapter 2,3,
(\bibinfo{publisher}{Princeton University Press},
\bibinfo{address}{Princeton}, \bibinfo{year}{1980}).

\bibitem[15]{Fill}
\bibinfo{author}
{\bibfnamefont{J.~A.}~\bibnamefont{Fillmore}}, 
\bibnamefont{and}
{\bibfnamefont{P.}~\bibnamefont{Goldreich}},
    \bibinfo{journal}{Astrophys.~J.} \textbf{\bibinfo{volume}{281}},
    \bibinfo{pages}{1} (\bibinfo{year}{1984}).

\bibitem[16]{Hoff1985}
\bibinfo{author}
{\bibfnamefont{Y.}~\bibnamefont{Hoffman}}, 
\bibnamefont{and}
{\bibfnamefont{J.}~\bibnamefont{Shaham}},
    \bibinfo{journal}{Astrophys.~J.} \textbf{\bibinfo{volume}{297}},
    \bibinfo{pages}{16} (\bibinfo{year}{1985}).

\bibitem[17]{Hoff1988}
\bibinfo{author}
{\bibfnamefont{Y.}~\bibnamefont{Hoffman}}, 
    \bibinfo{journal}{Astrophys.~J.} \textbf{\bibinfo{volume}{328}},
    \bibinfo{pages}{489} (\bibinfo{year}{1988}).

\bibitem[18]{Pers-1995}
\bibinfo{author}
{\bibfnamefont{M.}~\bibnamefont{Persic}},
\bibnamefont{and}
{\bibfnamefont{P.}~\bibnamefont{Salucci}},
    \bibinfo{journal}{Astron.~J.Suppl.~} \textbf{\bibinfo{volume}{99}},
    \bibinfo{pages}{501} (\bibinfo{year}{1995}).

\bibitem[19]{Pers-1996}
\bibinfo{author}
{\bibfnamefont{M.}~\bibnamefont{Persic}},
{\bibfnamefont{P.}~\bibnamefont{Salucci}},
\bibnamefont{and}
{\bibfnamefont{F.}~\bibnamefont{Stel}},
    \bibinfo{journal}{Mon.~Not.~R.~Astron.~Soc.} \textbf{\bibinfo{volume}{281}},
    \bibinfo{pages}{21} (\bibinfo{year}{1996}).

\bibitem[20]{Bender}
\bibinfo{author}{\bibfnamefont{C.~M.}
\bibnamefont{Bender}} \bibnamefont{and}
\bibinfo{author}{\bibfnamefont{S.~A.} \bibnamefont{Orszag}},
\emph{\bibinfo{title}{Advanced Mathematical Methods for Scientists and
    Engineers}}
(\bibinfo{publisher}{McGraw-Hill Book Company},
\bibinfo{address}{New York}, \bibinfo{year}{1978}).

\bibitem[21]{Davis2004}
\bibinfo{author}
{\bibfnamefont{T.}~\bibnamefont{Davis}},
\bibnamefont{and}
{\bibfnamefont{C.~H.}~\bibnamefont{Lineweaver}},
    \bibinfo{journal}{Publ.~Astron.~Soc.~Aust.~} \textbf{\bibinfo{volume}{21}},
    \bibinfo{pages}{97} (\bibinfo{year}{2004}).

\bibitem[22]{TurnerBook}
\bibinfo{author}{\bibfnamefont{E.~W.~}
\bibnamefont{Kolb}} \bibnamefont{and}
\bibinfo{author}{\bibfnamefont{M.~} \bibnamefont{Turner}},
\emph{\bibinfo{title}{The Early Universe}}, Chapter 9,
(\bibinfo{publisher}{Addison-Wesley Publishing Company},
\bibinfo{address}{New York}, \bibinfo{year}{1990}).

\bibitem[23]{McGa2005}
\bibinfo{author}
{\bibfnamefont{S.~S.}~\bibnamefont{McGaugh}},
    \bibinfo{journal}{Astron.~J.} \textbf{\bibinfo{volume}{632}},
    \bibinfo{pages}{859} (\bibinfo{year}{2005}).

\end{thebibliography}
\end{document}